\documentclass[aps,prc,preprint,groupedaddress,showpacs]{revtex4}
\usepackage[dvips]{graphicx}
\usepackage{amsmath,amssymb}

\begin{document}
\preprint{KUNS-1870}
\title{Nuclear moments of inertia and wobbling motions in triaxial superdeformed nuclei}

\author{Masayuki Matsuzaki}
\email[]{matsuza@fukuoka-edu.ac.jp}
\affiliation{Department of Physics, Fukuoka University of Education, 
             Munakata, Fukuoka 811-4192, Japan}

\author{Yoshifumi R. Shimizu}
\email[]{yrsh2scp@mbox.nc.kyushu-u.ac.jp}
\affiliation{Department of Physics, Graduate School of Sciences, Kyushu University,
             Fukuoka 812-8581, Japan}

\author{Kenichi Matsuyanagi}
\email[]{ken@ruby.scphys.kyoto-u.ac.jp}
\affiliation{Department of Physics, Graduate School of Science, Kyoto University,
             Kyoto 606-8502, Japan}

\date{\today}

\begin{abstract}
The wobbling motion excited on triaxial superdeformed nuclei is studied in terms of 
the cranked shell model plus random phase approximation. Firstly, by calculating 
at a low rotational frequency the $\gamma$-dependence of the three moments of inertia 
associated with the 
wobbling motion, the mechanism of the appearance of the wobbling motion in 
positive-$\gamma$ nuclei is clarified theoretically --- the rotational alignment 
of the $\pi i_{13/2}$ quasiparticle(s) is the essential condition. 
This indicates that the wobbling motion is a collective motion that is sensitive 
to the single-particle alignment. 
Secondly, we prove that the observed unexpected rotational-frequency dependence of the 
wobbling frequency is an outcome of the rotational-frequency dependent dynamical 
moments of inertia. 
\end{abstract}

\maketitle

\section{Introduction}

 Deformation of the nuclear shape from spherical symmetric one has long been 
one of the most important issues in nuclear structure physics. 
Among them, searches for evidences of the triaxial ($Y_{22}$ or $\gamma$) one 
have been pursued long time, for example, the even-odd energy staggering in the 
low-spin part of the $\gamma$ bands~\cite{davydov}, the signature dependence of 
the energy spectra and the $E2/M1$ transition rates in medium-spin odd-odd and odd-$A$ 
nuclei~\cite{bf,hm,mm2}, properties of the $K$ isomers~\cite{nari,tajima}, and so on. 
But their results have not been conclusive; making a clear distinction between the 
static and the dynamic (vibrational) ones has not been successful up to now. 
Theoretically, appearance of the wobbling motion, which is well-known in classical 
mechanics of asymmetric tops~\cite{landau} and whose quantum analog was discussed 
in terms of a rotor model about thirty years ago~\cite{bm}, is a decisive evidence 
of static triaxial deformations. Subsequently its microscopic descriptions were 
developed by several authors~\cite{jm,ma}. Since the small-amplitude wobbling 
mode carries the same quantum numbers, parity $\pi=+$ and signature $\alpha=1$, 
as the odd-spin members of the $\gamma$ band, Ref.\cite{mj} anticipated that it 
would appear as a high-spin continuation of the $\gamma$ band, but it has not 
been resolved that in what nuclei, at what spins, and with what $\gamma$ wobbling 
modes would be observed. 

 Shimizu and Matsuyanagi~\cite{sm} and Onishi~\cite{oni} performed extensive 
numerical calculations for normally-deformed Er isotopes with relatively small 
$|\gamma|$. Matsuzaki~\cite{mm}, Shimizu and Matsuzaki~\cite{smm}, and Horibata and 
Onishi~\cite{hori} also studied $^{182}$Os with relatively large negative-$\gamma$ 
but their correspondence to experimental information has not been very clear. 

 These studies indicate the necessity of high-spin states in stably and strongly 
$\gamma$-deformed nuclei. Bengtsson studied high-spin states around 
$^{164}$Hf~\cite{ragnar} and found systematic existence of the TSD (triaxial 
super- or strongly deformed) states with $\epsilon_2 \sim 0.4$ and 
$|\gamma| \sim 20^\circ$. This confirmed the discussion on the shell gap at $N=94$ in 
Ref.\cite{lu0}, the work in which the yrast TSD band in $^{163}$Lu was reported; and in 2000 
an excited TSD band was observed in this nucleus 
and from the strengths of the interband $E2$ transition rates this was unambiguously 
identified with the wobbling motion~\cite{lu1}. This data was analyzed by using a 
particle-rotor model~\cite{hama} and the $E2$ transition rates were reproduced well. 
Subsequently TSD bands were found in some Lu and Hf isotopes and 
wobbling excitations were observed also in $^{165,167}$Lu~\cite{lu165,lu167}. 
A close look at these data, however, tells us that the sign of their 
$\gamma$-deformation seems to contradict to an irrotational motion and that the 
unexpected behavior of the wobbling frequency has not been explained yet. 

 Thus in the preceding Rapid Communication~\cite{msmr} we presented an answer to 
these problems. In the present paper, after summarizing the discussion there we 
extend numerical analyses to elucidate it. An emphasis is put on the behavior of the 
calculated dynamic moments of inertia. 

\section{Wobbling motion in terms of the random phase approximation}
\label{sectwob}

 We start from a one-body Hamiltonian in the rotating frame, 
\begin{gather}
h'=h-\hbar\omega_\mathrm{rot}J_x , \\
h=h_\mathrm{Nil}-\mit\Delta_\tau (P_\tau^\dagger+P_\tau)
                   -\lambda_\tau N_\tau , \label{hsp} \\
h_\mathrm{Nil}=\frac{\mathbf{p}^2}{2M}
                +\frac{1}{2}M(\omega_x^2 x^2 + \omega_y^2 y^2 + \omega_z^2 z^2) \notag \\
                +v_{ls} \mathbf{l\cdot s} 
                +v_{ll} (\mathbf{l}^2 - \langle\mathbf{l}^2\rangle_{N_\mathrm{osc}}) .
                \label{hnil}
\end{gather}
In Eq.(\ref{hsp}), $\tau = 1$ and 2 stand for neutron and proton, respectively, 
and chemical potentials $\lambda_\tau$ are determined so as to give correct average 
particle numbers $\langle N_\tau \rangle$. 
The oscillator frequencies in Eq.(\ref{hnil}) 
are expressed by the quadrupole deformation parameters $\epsilon_2$ and $\gamma$ 
in the usual way. They are treated as parameters as well as pairing gaps 
$\mit\Delta_\tau$. 
The orbital angular momentum $\mathbf{l}$ in Eq.(\ref{hnil}) is defined in the 
singly-stretched coordinates $x_k' = \sqrt{\frac{\omega_k}{\omega_0}}x_k$, 
with $k =$ 1 -- 3 denoting $x$ -- $z$, and the corresponding momenta. 
By diagonalizing $h'$ at each $\omega_\mathrm{rot}$, we obtain quasiparticle (QP) 
orbitals and the nuclear yrast (0QP) state. 
Since $h'$ conserves parity $\pi$ and signature $\alpha$, nuclear states can be 
labeled by them. Nuclear states with QP excitations are obtained by exchanging 
the QP energy and wave functions such as 
\begin{equation}
(-e'_\mu, \mathbf{V}_\mu, \mathbf{U}_\mu) \rightarrow 
(e'_{\bar\mu}, \mathbf{U}_{\bar\mu}, \mathbf{V}_{\bar\mu}) ,
\label{exch}
\end{equation}
where ${\bar\mu}$ denotes the signature partner of $\mu$. 

 We perform the random phase approximation (RPA) to the residual pairing plus 
doubly-stretched quadrupole-quadrupole ($Q'' \cdot Q''$) interaction between QPs. 
Since we are interested in the wobbling motion that has a definite quantum number, 
$\alpha = 1$, only two components out of five of the $Q'' \cdot Q''$ interaction 
are relevant. They are given by 
\begin{equation}
H_\mathrm{int}^{(-)}=-\frac{1}{2}\sum_{K=1,2} \kappa_K^{(-)} Q_K''^{(-)\dagger} Q_K''^{(-)} ,
\end{equation}
where the doubly-stretched quadrupole operators are defined by 
\begin{equation}
Q_K''=Q_K(x_k\rightarrow x_k'' = \frac{\omega_k}{\omega_0}x_k) ,
\end{equation}
and those with good signature are 
\begin{equation}
Q_K^{(\pm)}=\frac{1}{\sqrt{2(1+\delta_{K0})}}\left(Q_K \pm Q_{-K}\right) .
\end{equation}
The residual pairing interaction does not contribute because $P_\tau$ is an 
operator with $\alpha = 0$. 
The equation of motion, 
\begin{equation}
\left[h'+H_\mathrm{int}^{(-)},X_n^\dagger\right]_\mathrm{RPA}
=\hbar\omega_n X_n^\dagger ,
\end{equation}
for the eigenmode 
\begin{equation}
X_n^\dagger=\sum_{\mu<\nu}^{(\alpha=\pm 1/2)}
\Big(\psi_n(\mu\nu)a_\mu^\dagger a_\nu^\dagger 
+\varphi_n(\mu\nu)a_\nu a_\mu\Big)
\end{equation}
leads to a pair of coupled equations for the transition amplitudes 
\begin{equation}
T_{K,n}=\left\langle\left[Q_K^{(-)},X_n^\dagger\right]\right\rangle .
\end{equation}
Then, by assuming $\gamma \neq 0$, this can be cast~\cite{ma} into the form 
\begin{gather}
(\omega_n^2-\omega_\mathrm{rot}^2)\left[\omega_n^2-\omega_\mathrm{rot}^2
\frac{\left(\mathcal{J}_x-\mathcal{J}_y^\mathrm{(eff)}(\omega_n)\right)
   \left(\mathcal{J}_x-\mathcal{J}_z^\mathrm{(eff)}(\omega_n)\right)}
{\mathcal{J}_y^\mathrm{(eff)}(\omega_n)\mathcal{J}_z^\mathrm{(eff)}(\omega_n)}
                                  \right] \notag \\
=0 ,
\end{gather}
which is independent of $\kappa_K^{(-)}$s. 
This expression proves that the spurious (Nambu--Goldstone) mode given by the first 
factor and all nomal modes given by the second are decoupled from each other. 
Here $\mathcal{J}_x = \langle J_x \rangle/\omega_\mathrm{rot}$ as usual and the 
detailed expressions of $\mathcal{J}_{y,z}^\mathrm{(eff)}(\omega_n)$ are given in 
Refs.\cite{ma,mm,smm}. Among normal modes, one obtains 
\begin{equation}
\omega_\mathrm{wob}^2=\omega_\mathrm{rot}^2
\frac{\left(\mathcal{J}_x-\mathcal{J}_y^\mathrm{(eff)}(\omega_\mathrm{wob})\right)
   \left(\mathcal{J}_x-\mathcal{J}_z^\mathrm{(eff)}(\omega_\mathrm{wob})\right)}
     {\mathcal{J}_y^\mathrm{(eff)}(\omega_\mathrm{wob})
      \mathcal{J}_z^\mathrm{(eff)}(\omega_\mathrm{wob})} ,
\label{disp}
\end{equation}
by putting $\omega_n=\omega_\mathrm{wob}$. 
Note that this gives a real excitation only when the right-hand side is positive 
and it is non-trivial whether a collective solution appears or not. 
Evidently this coincides with the form derived by Bohr and Mottelson in a rotor 
model~\cite{bm} and known in classical mechanics~\cite{landau}, aside from the 
crucial feature that the moments of inertia are $\omega_\mathrm{rot}$-dependent in the 
present case. 

 One drawback in our formulation is that our $\mathcal{J}_x$ tends to be larger than 
corresponding experimental values because of the spurious velocity dependence of 
the Nilsson potential as discussed in Refs.\cite{kino,chun}. 
A remedy for this was discussed 
there but that for $\mathcal{J}_{y,z}^\mathrm{(eff)}$ has not been devised yet. 
Therefore we assume for the present a similar discussion holds for the latter and 
accordingly the ratio 
$\mathcal{J}_{y,z}^\mathrm{(eff)}(\omega_\mathrm{wob})/\mathcal{J}_x$ which 
actually determines $\omega_\mathrm{wob}$ is more reliable than their absolute 
magnitudes. 

 Interband electric quadrupole transitions between the $n$-th excited band and the 
yrast are given as
\begin{equation}
B(E2:I_n \rightarrow (I \pm 1)_\mathrm{yrast})
=\frac{1}{2}\left(T_{1,n}^{(E)} \pm T_{2,n}^{(E)}\right)^2 ,
\end{equation}
in terms of 
\begin{equation}
T_{K,n}^{(E)}=e\frac{Z}{A}T_{K,n} .
\end{equation}
They will be abbreviated to $B(E2)_\mathrm{out}$ later for simplicity. 
In-band ones are given as
\begin{equation}
B(E2:I \rightarrow I-2)
=\frac{1}{2}\left(\frac{\sqrt{3}}{2}\left\langle Q_0^{(+)(E)}\right\rangle 
         + \frac{1}{2}\left\langle Q_2^{(+)(E)}\right\rangle\right)^2 ,
\end{equation}
in terms of 
\begin{equation}
\left\langle Q_K^{(+)(E)}\right\rangle
=e\frac{Z}{A}\left\langle Q_K^{(+)}\right\rangle ,
\end{equation}
and assumed to be common to all bands. They will be abbreviated to $B(E2)_\mathrm{in}$. 
Here we adopted a high-spin approximation~\cite{ma2}. 
The transition quadrupole moment $Q_\mathrm{t}$ is extracted from $B(E2)_\mathrm{in}$ 
by the usual rotor-model prescription. 

 To compare collectivities of these two types of $E2$ transitions, we introduce a 
pair of deformation parameters
\begin{gather}
R^2\alpha_y=\sqrt{\frac{15}{16\pi}}\left\langle x^2-z^2 \right\rangle
 =\left\langle\frac{1}{2}Q_2^{(+)}-\frac{\sqrt{3}}{2}Q_0^{(+)}\right\rangle, \notag \\
R^2\alpha_z=\sqrt{\frac{15}{16\pi}}\left\langle x^2-y^2 \right\rangle
 =\left\langle Q_2^{(+)}\right\rangle.
\end{gather}
Then it is evident that the in-band one is expressed as
\begin{equation}
B(E2:I \rightarrow I-2)
=\frac{1}{2}R^4\,\Big(\alpha_y^{(E)}-\alpha_z^{(E)}\Big)^2.
\end{equation}
As for the interband ones, by expanding $Q_K^{(-)}$ by $X_n^\dagger$s and $X_n$s, 
where $n$ runs both normal modes and the Nambu--Goldstone mode 
$X_\mathrm{NG}^\dagger=\frac{1}{\sqrt{2I}}(J_z+iJ_y)$, we obtain from 
$[Q_1^{(-)},Q_2^{(-)}]=0$ a kind of sum rule 
\begin{equation}
\sum_{n\neq\mathrm{NG}}T_{1,n}T_{2,n}=-\frac{2}{I}\,R^4\,\alpha_y\alpha_z.
\end{equation}
Consecutively introducing the ratios of the dynamic to static deformations, 
\begin{gather}
r_{y,n}=\frac{T_{1,n}}{2R^2\alpha_y}, \notag \\
r_{z,n}=-\frac{T_{2,n}}{2R^2\alpha_z},
\end{gather}
the sum rule above reads
\begin{equation}
\sum_{n\neq\mathrm{NG}}r_{y,n}r_{z,n}=\frac{1}{2I}.
\label{sumrl}
\end{equation}
The dynamic amplitudes $T_{K,n}$ describe shape fluctuations associated
with the vibrational motion in the uniformly rotating frame.
Transformation to the body-fixed (Principal-Axis) frame~\cite{ma} turns
the shape fluctuation into the fluctuation of the angular momentum vector,
i.e., the wobbling motion.  This transformation relates the ratios,
$r_{y,n}$ and $r_{z,n}$, to the moments of inertia~\cite{smm}:
\begin{gather}
r_{y,n}=c_n\frac{1}{\sqrt{2I}}\left(\frac{W_{z,n}}{W_{y,n}}\right)^{1/4}, \notag \\
r_{z,n}=\sigma_nc_n\frac{1}{\sqrt{2I}}\left(\frac{W_{y,n}}{W_{z,n}}\right)^{1/4},
\end{gather}
where $c_n$ is a real amplitude that relates the dynamic amplitude
$T_{K,n}$ and the moment of inertia, $\sigma_n$ is
the sign of $\left(\mathcal{J}_x-\mathcal{J}_y^\mathrm{(eff)}\right)$
(so $\sigma_n > 0$ for wobbling-like RPA solutions), and 
\begin{gather}
W_{y,n}=1/\mathcal{J}_z^\mathrm{(eff)}(\omega_n) -1/\mathcal{J}_x, \notag \\
W_{z,n}=1/\mathcal{J}_y^\mathrm{(eff)}(\omega_n) -1/\mathcal{J}_x.
\end{gather}
Thus, the interband $B(E2)$ is rewritten as,
\begin{gather}
B(E2:I_n \rightarrow (I \pm 1)_\mathrm{yrast}) \notag \\
=\frac{1}{I}\,R^4\,c_n^2\,\left[
\alpha_y^{(E)}\left(\frac{W_{z,n}}{W_{y,n}}\right)^{1/4} \mp \sigma_n
\alpha_z^{(E)}\left(\frac{W_{y,n}}{W_{z,n}}\right)^{1/4}\right]^2,
\label{BE2wob}
\end{gather}
which coincides with the formula given by the rotor model~\cite{bm},
except for the appearance of the amplitude $c_n$ and sign $\sigma_n$.
Substituting the ratios, $r_{y,n}$ and $r_{z,n}$, into Eq.(\ref{sumrl}),
one finds that the amplitudes should satisfy
\begin{equation}
\sum_{n\neq\mathrm{NG}}\sigma_n c_n^2=1.
\label{sum}
\end{equation}
This form of sum rule clearly indicates that the amplitude $c_n$
is a microscopic correction factor quantifying the collectivity
of the wobbling motion, for which
$c_n^2 \simeq 1$ means the full-collectivity and reproduces
the results of the macroscopic rotor model in both the energy
and the interband $B(E2)$ values.

\section{Numerical calculation and Discussion}

\subsection{Summary of the preceding study}

 Since the first experimental confirmation of the wobbling excitation in 
$^{163}$Lu~\cite{lu1}, $\gamma \simeq +20^\circ$ has been widely accepted as the shape 
of the TSD states in this region. This is predominantly because the calculated 
energy minimum for $\gamma \simeq +20^\circ$ is deeper than that for 
$\gamma \simeq -20^\circ$~\cite{ragnar} 
according to the shape driving effect of the aligned $\pi i_{13/2}$ quasiparticle. 
The recent precise measurements of $Q_\mathrm{t}$~\cite{luqt} also support this. 
On the other hand, the sign of $\gamma$-deformation leads
different consequences on moments of inertia, which are directly connected
to the excitation energy of the wobbling mode through
the wobbling frequency formula~\cite{bm}, c.f. Eq.(\ref{disp}).
Since the RPA is a microscopic formalism, no distinction between the collective rotation 
and the single-particle degrees of freedom has been made.

Therefore, the moments of inertia calculated in our RPA formalism in sect.\ref{sectwob}
are those for rotational motions of the whole system.
In contrast, the macroscopic irrotational-like moments of inertia are
often used in the particle-rotor calculations, where
$\mathcal{J}_y > \mathcal{J}_x \gg \mathcal{J}_z$ for $\gamma \simeq +20^\circ$ 
and they lead to an imaginary wobbling frequency $\omega_\mathrm{wob}$.
It is, however, noted that the moments of inertia of the particle-rotor model
are those of the rotor and no effect of the single-particle alignments is
included, so that they do not necessarily correspond to those
calculated in our RPA formalism.

 In the preceding paper~\cite{msmr} we have performed microscopic RPA
calculations without dividing the system artificially into the rotor and
particles.  That work proved that for the calculated moment of inertia,
$\mathcal{J}_x = \langle J_x \rangle/\omega_\mathrm{rot}$,
the contribution from the aligned QP(s),
$\mit\Delta\mathcal{J}_x = i_\mathrm{QP}/\omega_\mathrm{rot}$ with 
$i_\mathrm{QP}$ being the aligned angular momentum, is superimposed 
on an irrotational-like moment of inertia ($\mathcal{J}_y > \mathcal{J}_x$)
of the ``core".  Consequently
the total $\mathcal{J}_x$ is larger than $\mathcal{J}_y$,
which makes wobbling excitation in $\gamma > 0$ nuclei possible.

 The second consequence of the formulation adopted in Ref.\cite{msmr} is that the 
three moments of inertia are automatically $\omega_\mathrm{rot}$-dependent even 
when the mean-field parameters are fixed constant. This is essential in order to 
explain the observed $\omega_\mathrm{rot}$-dependence of $\omega_\mathrm{wob}$ 
--- \textit{decreasing} as $\omega_\mathrm{rot}$ increases. 
Otherwise $\omega_\mathrm{wob}$ is proportional to $\omega_\mathrm{rot}$. 

 Another important feature of the data is that the interband $B(E2)$ values between the wobbling 
and the yrast TSD bands are surprisingly large. Our RPA wave function gave extremely 
collective $B(E2)_\mathrm{out}$
that gathered $|c_{n=\mathrm{wob}}| \simeq$ 0.6 -- 0.8 
in the sum rule (Eq.(\ref{sum})) 
but the result accounted for only about one half of the measured one. 

 To elucidate these findings more, in the following we extend our numerical 
analyses putting a special emphasis on the $\gamma$-dependence of the moments of inertia 
in subsect.\ref{sectlow}. 
Dependence on other parameters is also studied in detail. 
Features in common and different between even-even and odd-$A$ nuclei are 
also pointed out. 
In subsect.\ref{secthigh}, we discuss $\omega_\mathrm{rot}$-dependence. 
In subsect.\ref{sectbe2}, characteristics of $B(E2)_\mathrm{out}$ are discussed. 
Calculations are performed in five major shells; $N_\mathrm{osc} =$ 3 -- 7 for 
neutrons and $N_\mathrm{osc} =$ 2 -- 6 for protons. The strengths $v_{ls}$ and 
$v_{ll}$ in Eq.(\ref{hnil}) are taken from Ref.\cite{br}. 

\subsection{Dependence on the mean-field parameters $\gamma$, $\epsilon_2$, and $\mit\Delta$
}
\label{sectlow}

\subsubsection{The even-even nucleus $^{168}$Hf}
\label{sectlowHf}

 Hafnium-168 is the first even-even nucleus in which TSD bands were 
observed~\cite{hf168}. In this nucleus three TSD bands 
were observed but interband $\gamma$-rays connecting them have not been observed yet. 
This means that the character of the excited bands has not been established, 
although we expect at least one of them is wobbling excitation. 
An important feature of the data is that the average transition quadrupole 
moment was determined as $Q_\mathrm{t} = 11.4^{+1.1}_{-1.2}$ eb. 
This imposes a moderate constraint on the shape. Referring to 
the weak parameter dependence discussed later, we choose 
$\epsilon_2 = 0.43$, $\gamma = 20^\circ$, and $\mit\Delta_n = \mit\Delta_p =$ 0.3 MeV, 
which reproduce the observed $Q_\mathrm{t}$, as a typical mean-field parameter set. 

 First we study the dependence of various quantities on $\gamma$ and other mean-field 
parameters at $\hbar\omega_\mathrm{rot} =$ 0.25 MeV. Around this frequency the 
$(\pi i_{13/2})^2$ alignment that is essential for making wobbling excitation in 
$\gamma > 0$ nuclei possible is completed and therefore the wobbling motion 
is expected to emerge above this frequency (see Fig.\ref{fig7} shown later). 

\begin{figure}[htbp]
  \includegraphics[width=7cm,keepaspectratio]{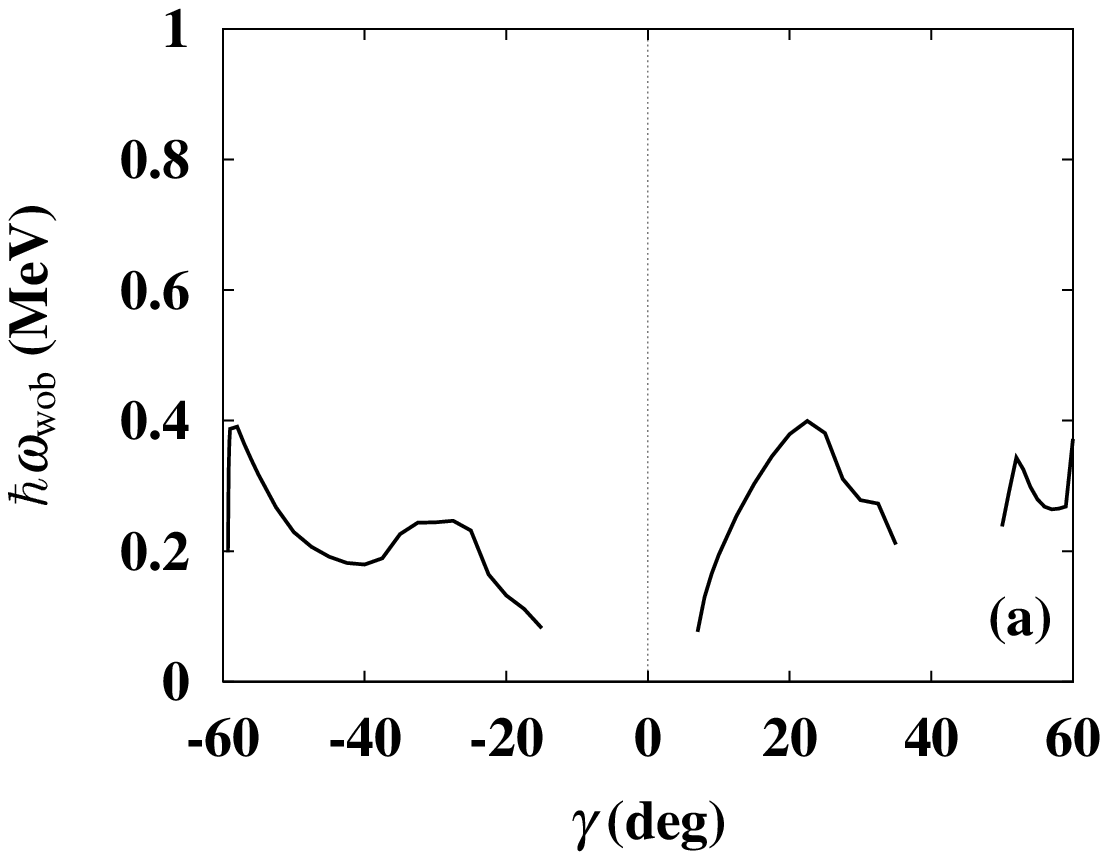}
  \includegraphics[width=7cm,keepaspectratio]{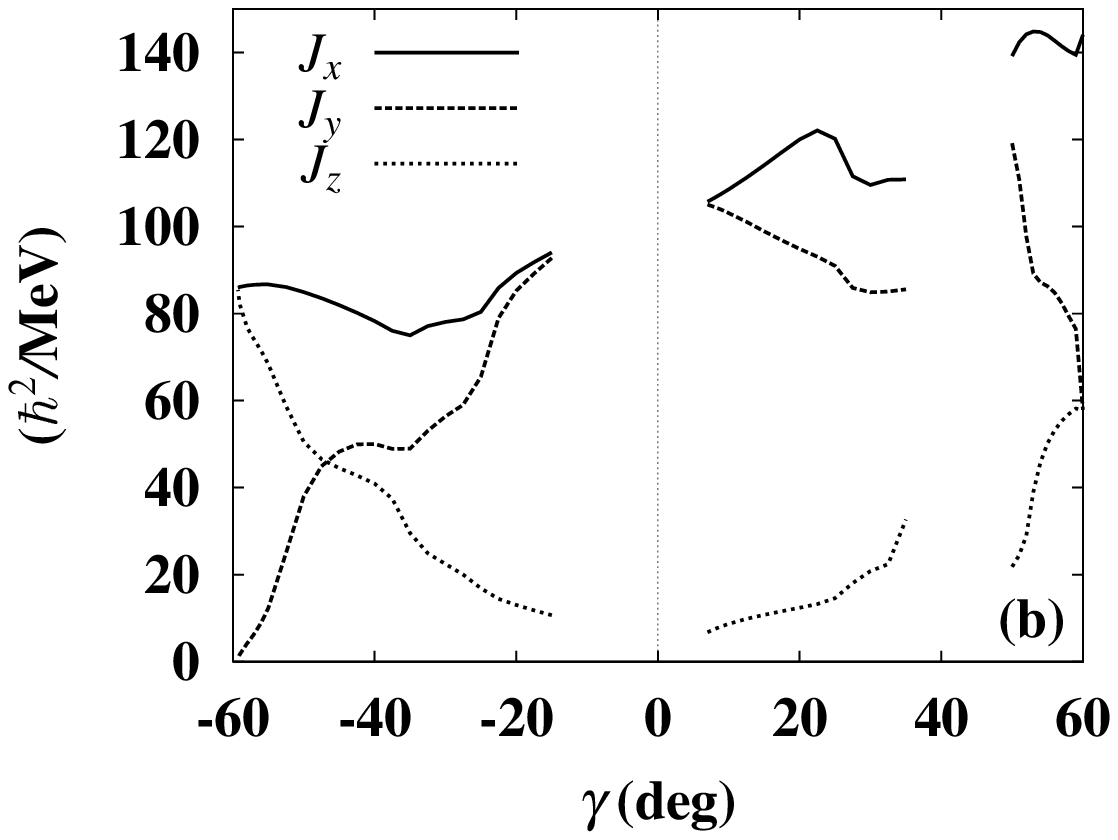}
  \includegraphics[width=7cm,keepaspectratio]{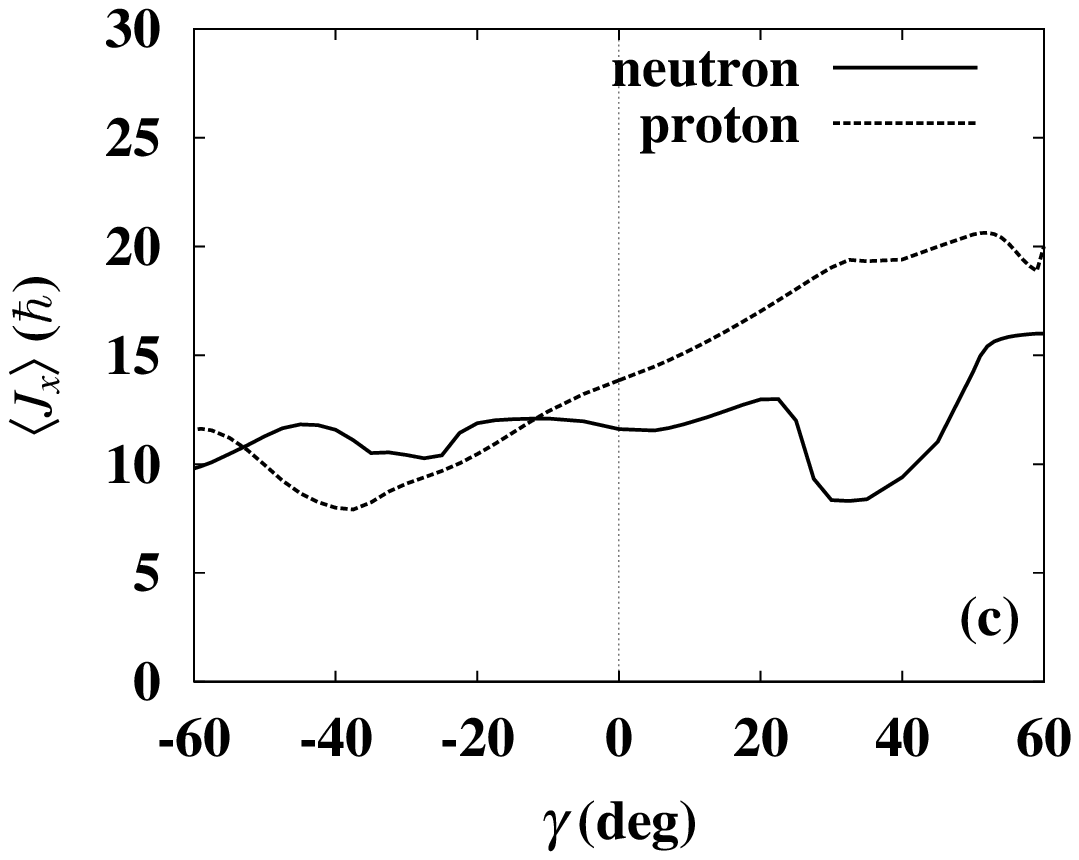}
  \includegraphics[width=7cm,keepaspectratio]{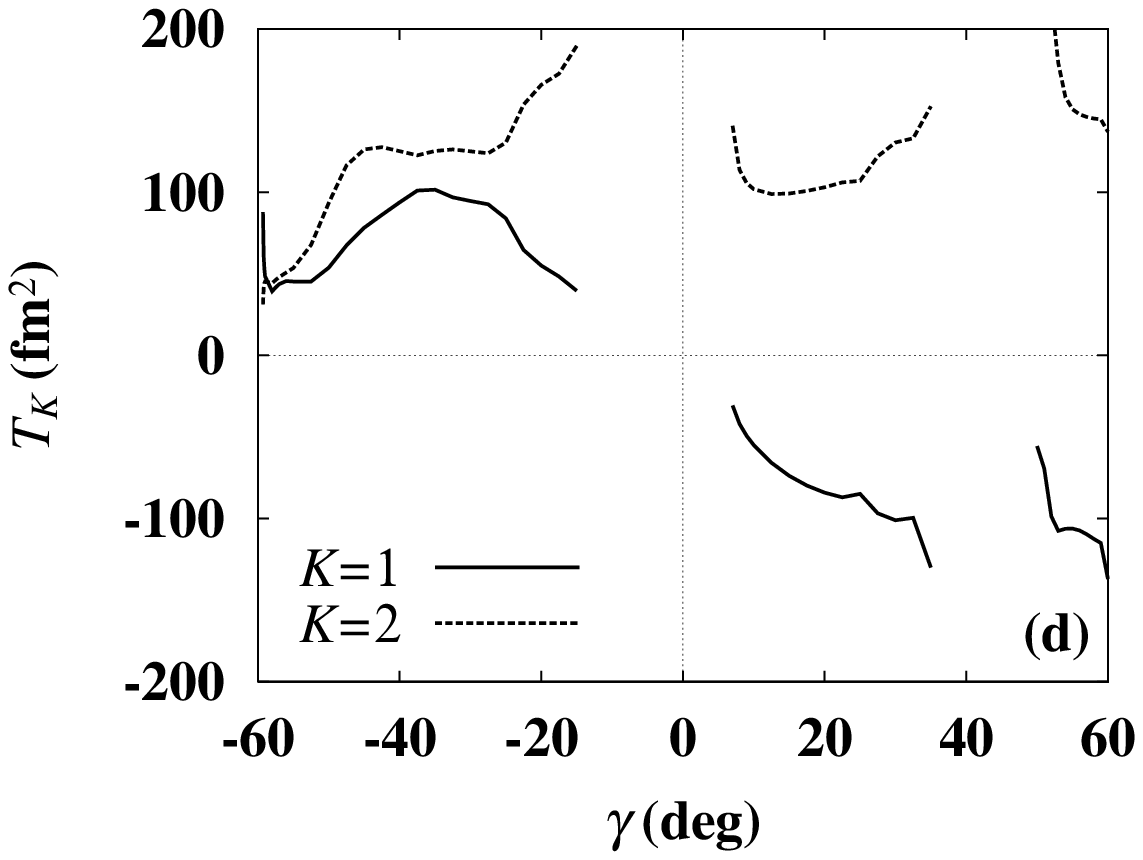}
 \caption{Triaxiality dependence of (a) excitation energy of the wobbling motion, 
          (b) three moments of inertia associated with it, (c) expectation values 
          of angular momenta in the yrast state, and (d) quadrupole transition 
          amplitudes between the wobbling and the yrast states in $^{168}$Hf, 
          calculated at $\hbar\omega_\mathrm{rot} =$ 0.25 MeV with 
          $\epsilon_2 = 0.43$ and $\mit\Delta_n = \mit\Delta_p =$ 0.3 MeV. \label{fig1}}
\end{figure}

 Figure~\ref{fig1} shows dependence on $\gamma$ calculated with keeping 
$\epsilon_2 = 0.43$ and $\mit\Delta_n = \mit\Delta_p =$ 0.3 MeV. Figure~\ref{fig1}(a) graphs 
the calculated excitation energy in the rotating frame, $\hbar\omega_\mathrm{wob}$. 
As $\gamma$ comes close to 0 (symmetric about the $z$ axis) and $-60^\circ$ 
(symmetric about the $y$ axis), $\omega_\mathrm{wob}$ approaches 0, see Eq.(\ref{disp}). 
We did not obtain any low-lying RPA solutions at around $\gamma = 40^\circ$ 
whereas a collective solution appears again for $50^\circ \le \gamma \le 60^\circ$. 

 Figure~\ref{fig1}(b) shows the calculated moments of inertia. Their 
$\gamma$-dependence resembles the irrotational, the so-called $\gamma$-reversed, and 
the rigid-body moments of inertia, in $\gamma <  0$, $0 < \gamma < 40^\circ$, 
and $50^\circ \le \gamma \le 60^\circ$, respectively. 
These model moments of inertia are given by 
\begin{equation}
\mathcal{J}_k^\mathrm{irr}=4B\beta^2\sin^2{(\gamma+\frac{2}{3}\pi k)} ,
\label{irr}
\end{equation}
\begin{equation}
\mathcal{J}_k^\mathrm{rev}=4B\beta^2\sin^2{(-\gamma+\frac{2}{3}\pi k)} ,
\label{rev}
\end{equation}
and
\begin{equation}
\mathcal{J}_k^\mathrm{rig}=\mathcal{J}_0
\left(1-\sqrt{\frac{5}{4\pi}}\beta\cos{(\gamma+\frac{2}{3}\pi k)}\right) ,
\label{rig}
\end{equation}
where $k =$ 1 -- 3 denote the $x$ -- $z$ principal axes, $B$ the irrotational 
mass parameter, $\mathcal{J}_0$ the rigid moment of inertia in the spherical 
limit, and $\beta$ is a deformation parameter like $\epsilon_2$. 
The $\gamma$-reversed moment of inertia was introduced to describe positive-$\gamma$ 
rotations in the particle-rotor model~\cite{hm} but its physical meaning has not 
been very clear; in particular, it does not fulfill the quantum-mechanical 
requirement that the rotations about the symmetry axis should be forbidden. 
We have clarified in the preceding paper~\cite{msmr} that the contributions from 
aligned quasiparticles superimposed on irrotational-like moments of inertia 
($\mathcal{J}_x < \mathcal{J}_y$) can realize $\mathcal{J}_x > \mathcal{J}_y$ 
and this is the very reason why the wobbling excitation (see Eq.(\ref{disp})) 
appears in positive-$\gamma$ nuclei. We also discussed that multiple alignments 
could eventually lead to a rigid-body-like moment of inertia. 
Figure~\ref{fig1}(c) indicates that, in the present calculation in which 
configuration is specified as the adiabatic quasiparticle vacuum at each 
$\omega_\mathrm{rot}$, two $\pi i_{13/2}$ protons align for $\gamma > 0$ as mentioned 
above while they have not fully aligned for $\gamma < 0$ at this $\omega_\mathrm{rot}$. 
In other words, these figures cover the both regions in which the $(\pi i_{13/2})^2$ 
alignment is necessary ($\gamma>0$) and that is not necessary ($\gamma<0$) for obtaining 
wobbling excitations. This aligned angular momentum determines the overall 
$\gamma$-dependence of $\mathcal{J}_x$ in Fig.\ref{fig1}(b). 
As for the neutron part, corresponding to the disappearance of the solution at 
around $\gamma = 40^\circ$, the expectation value of the neutron angular momentum, 
$\langle J_x\rangle_n$, drops around this region. 

\begin{figure}[htbp]
  \includegraphics[width=7cm,keepaspectratio]{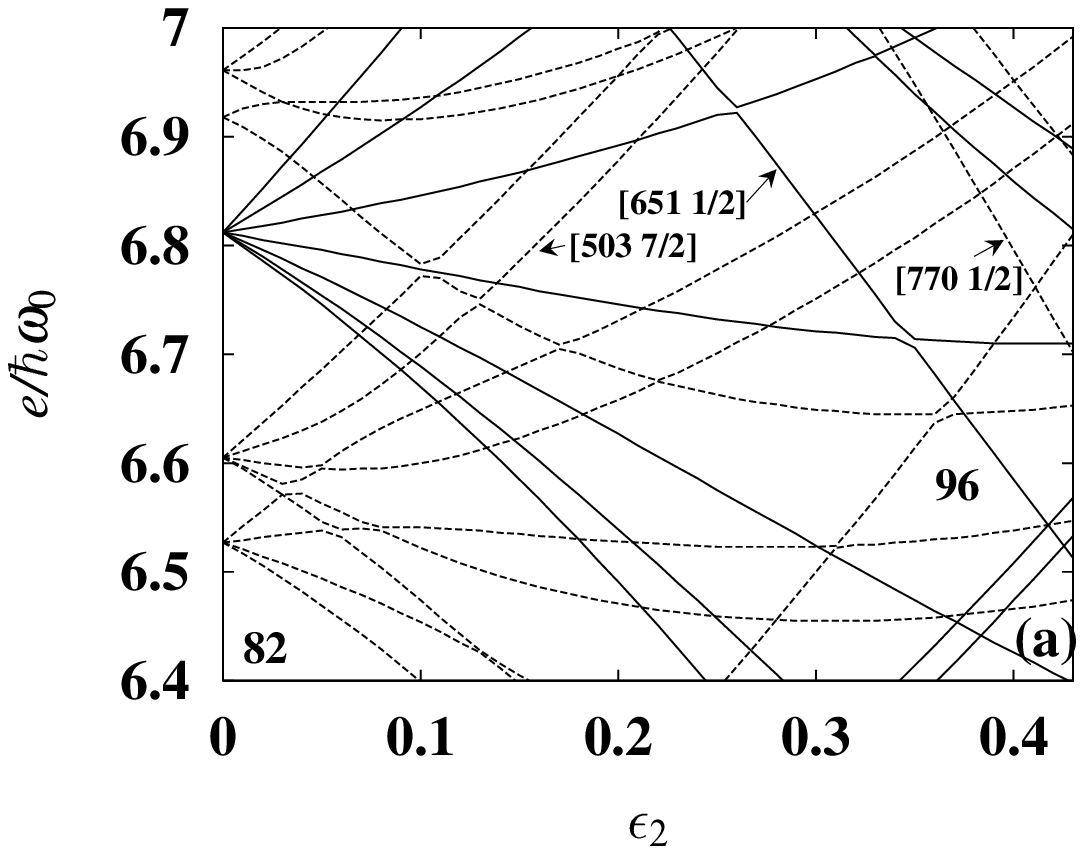}
  \includegraphics[width=7cm,keepaspectratio]{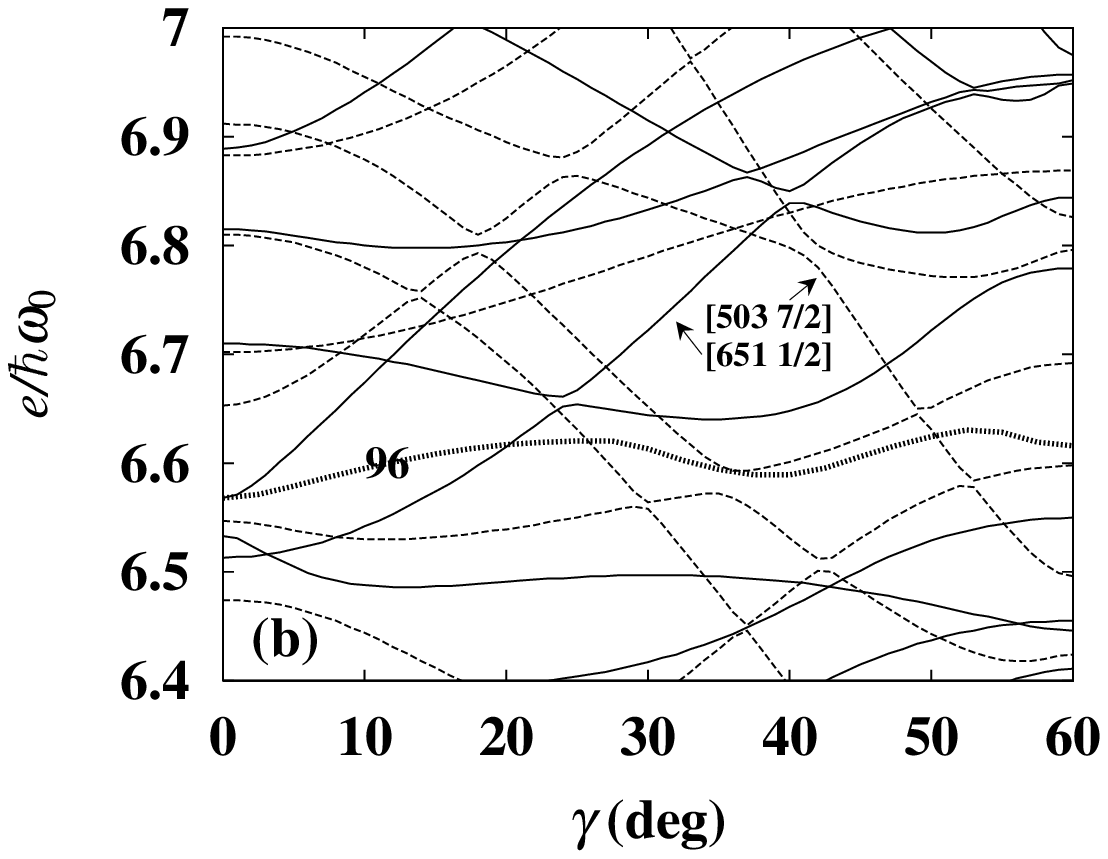}
  \includegraphics[width=7cm,keepaspectratio]{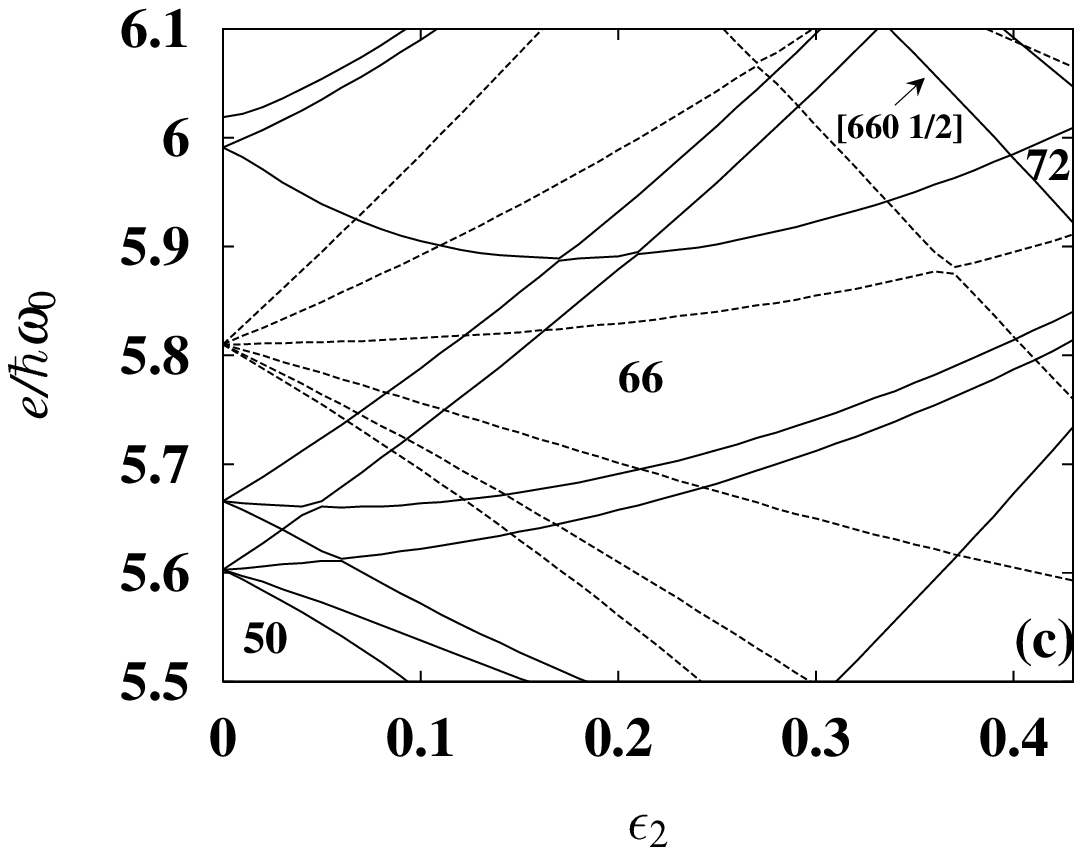}
  \includegraphics[width=7cm,keepaspectratio]{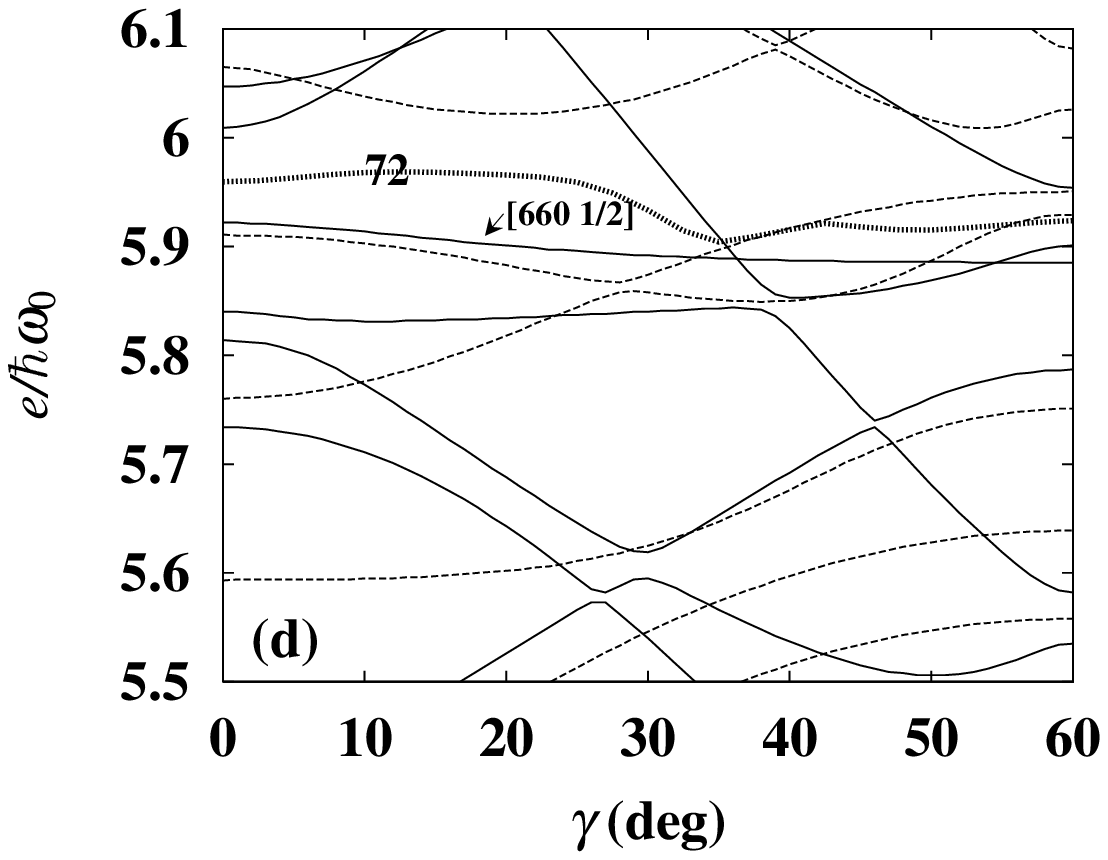}
 \caption{Nilsson single-particle energy diagrams at $\omega_\mathrm{rot} = 0$, 
          (a) for $0 \le \epsilon_2 \le 0.43$ with $\gamma = 0$, and (b) for 
          $0 \le \gamma \le 60^\circ$ with $\epsilon_2 = 0.43$ for neutrons. 
          (c) and (d) are corresponding ones for protons. 
          Solid and dashed curves represent even and odd parity orbitals, 
          respectively. Asymptotic quantum numbers of some important orbitals 
          are explicitly indicated. Chemical potentials that give particle numbers 
          $N = 96$ and $Z = 72$ for $\gamma = +20^\circ$  at 
          $\hbar\omega_\mathrm{rot}$ = 0.25 MeV are also indicated in (b) and (d). 
          \label{fig2}}
\end{figure}

 To look at this more closely, we investigate the Nilsson single-particle diagram at 
$\omega_\mathrm{rot} = 0$. Figure~\ref{fig2}(a) graphs neutron single-particle 
energies for $0 \le \epsilon_2 \le 0.43$ with $\gamma = 0$, while Fig.\ref{fig2}(b) 
for $0 \le \gamma \le 60^\circ$ with $\epsilon_2 = 0.43$. 
The chemical potential that gives correct neutron number $N = 96$ for $\gamma > 0$ 
at $\hbar\omega_\mathrm{rot}$ = 0.25 MeV is also drawn in the latter. 
This figure clearly shows that with this $\epsilon_2$ a shell gap exists for 
$\gamma \alt 20^\circ$ at $N = 96$. 
And by comparing this with Fig.\ref{fig1} we see that the dropping of 
$\langle J_x\rangle_n$ is a consequence of the deoccupation of 
the orbital that is [651~1/2] at $\gamma = 0$ (hereafter simply referred to as the [651~1/2] 
orbital even at $\gamma \neq 0$) originating from the mixed ($g_{9/2}$-$i_{11/2}$) spherical shell. 
Figure~\ref{fig2}(b) also explains the reason why the wobbling excitation revives 
at around $\gamma = 50^\circ$ again; the occupation of other oblate-favoring orbitals such 
as [503~7/2] makes it possible and leads to a rigid-body-like behavior of the moments 
of inertia. Figures~\ref{fig2}(c) and (d) are corresponding ones 
for protons. This indicates that the proton shell gap is robuster. 

 Figure~\ref{fig1}(d) graphs the quadrupole transition amplitudes 
$T_K$ ($K=1,2$) associated with the wobbling mode. ($T_K$ corresponds to 
$(-1)^{K-1}\mathcal{Q}_K$ in Ref.\cite{smm}.)  This shows that their relative sign 
changes with that of $\gamma$ as discussed in Refs.\cite{mm,smm}. 
This feature can be understood as: 
$\gamma \sim 0$ is the $\gamma$-vibrational 
region because the $K = 2$ component is dominant (see also 
$\mathcal{J}_x \simeq \mathcal{J}_y^\mathrm{(eff)}$ and 
$\mathcal{J}_z^\mathrm{(eff)}  \simeq 0$ in Fig.\ref{fig1}(b)), and 
the mixing of the $K = 1$ component due to triaxiality and rotation gives rise to the 
character of the wobbling motion. This relative sign leads to a selection rule 
of the interband transition probabilities $B(E2)_\mathrm{out}$~\cite{smm}. 
In the present case we obtain 
$B(E2:I \rightarrow I-1)_\mathrm{out} \gtrless B(E2:I \rightarrow I+1)_\mathrm{out}$ 
for $\gamma \gtrless 0$, and typically their ratio to the in-band ones is 
$B(E2:I \rightarrow I-1)_\mathrm{out} / B(E2:I \rightarrow I-2)_\mathrm{in} \sim 0.1$.

\begin{figure}[htbp]
  \includegraphics[width=7cm,keepaspectratio]{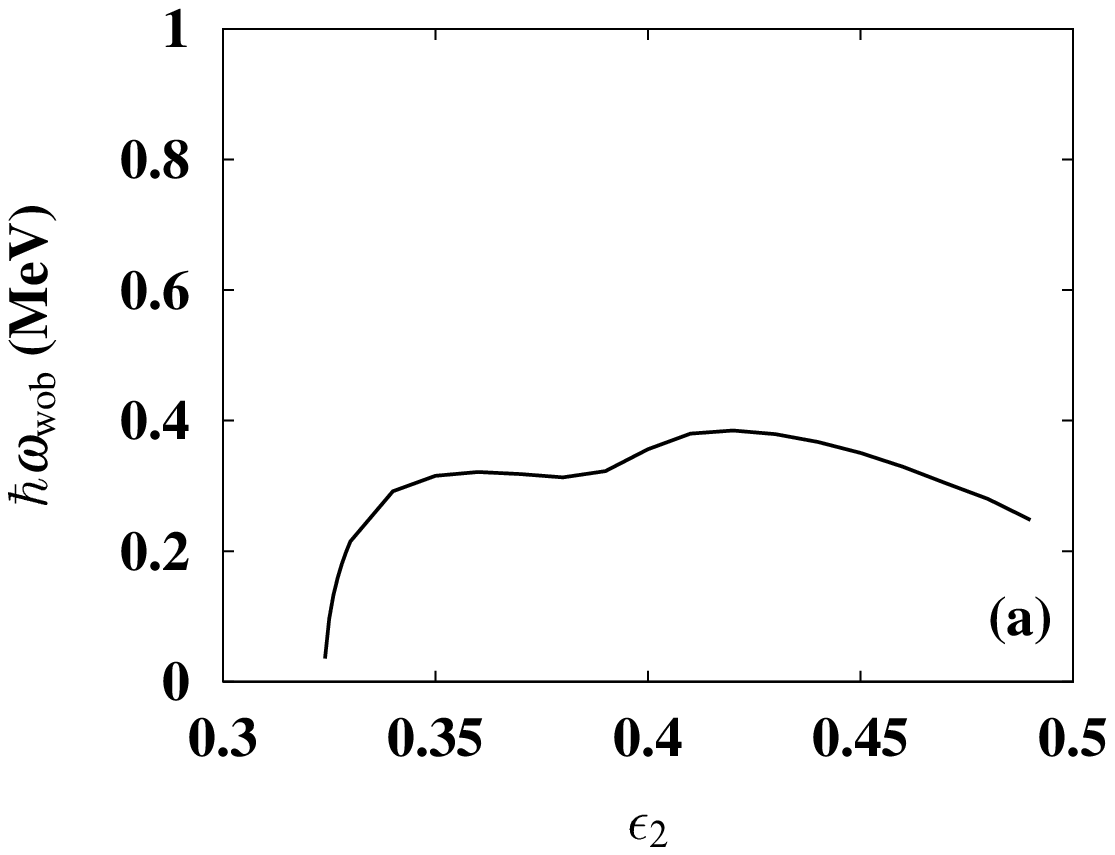}
  \includegraphics[width=7cm,keepaspectratio]{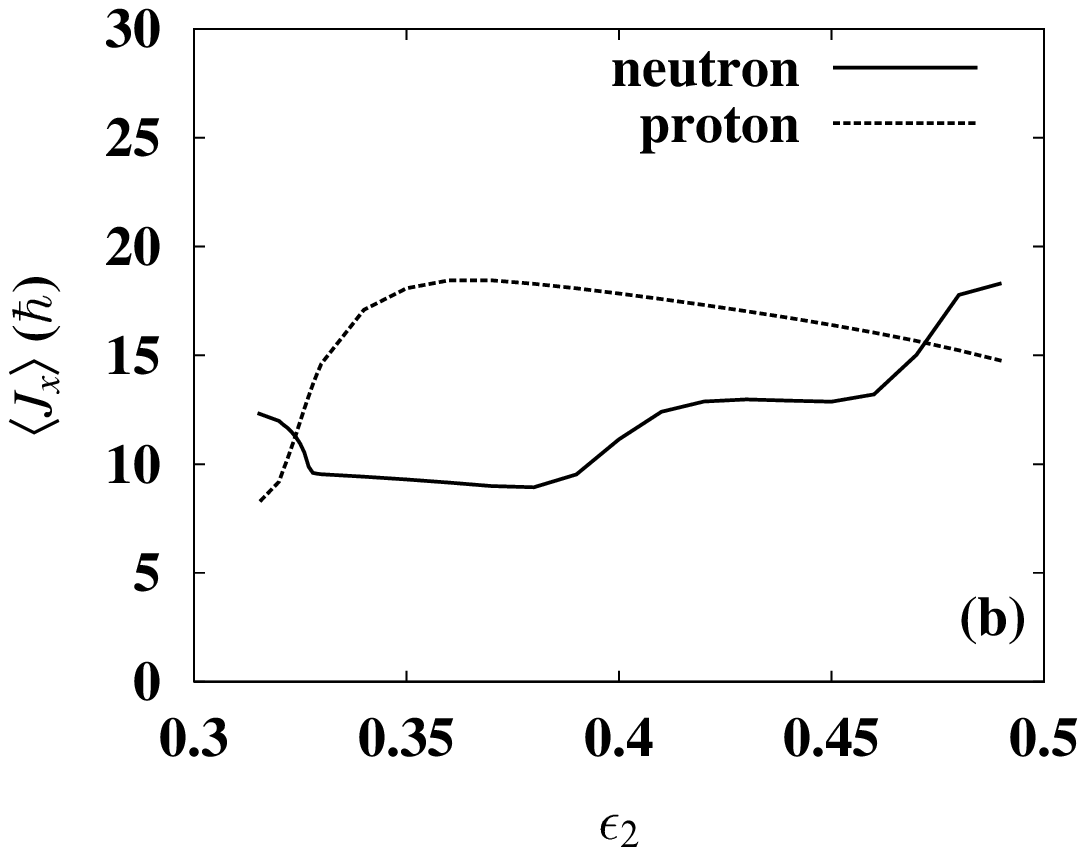}
  \includegraphics[width=7cm,keepaspectratio]{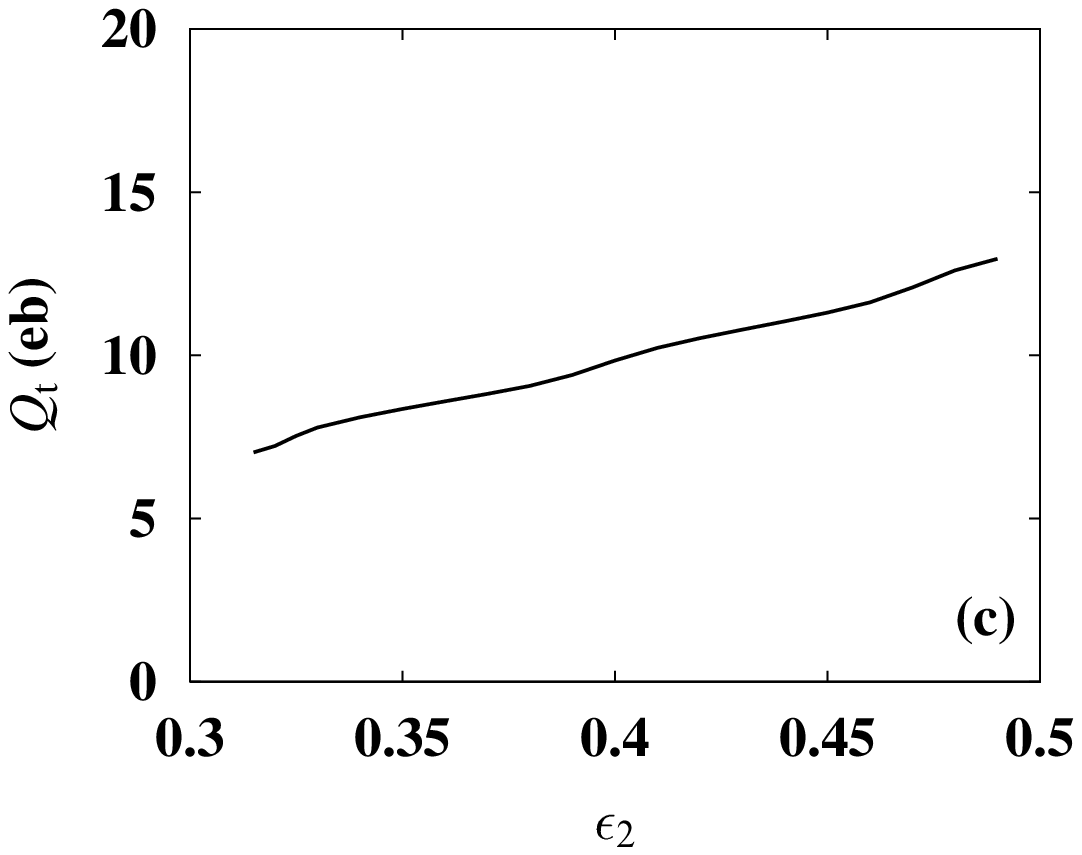}
 \caption{Deformation dependence of (a) excitation energy of the wobbling motion, 
          (b) expectation values of angular momenta in the yrast state, and 
          (c) transition quadrupole moment in the yrast state in $^{168}$Hf, 
          calculated at $\hbar\omega_\mathrm{rot} =$ 0.25 MeV with 
          $\gamma = 20^\circ$ and $\mit\Delta_n = \mit\Delta_p =$ 0.3 MeV. \label{fig3}}
\end{figure}

 Figure~\ref{fig3} shows dependence on $\epsilon_2$ calculated with keeping 
$\gamma = 20^\circ$ and $\mit\Delta_n = \mit\Delta_p =$ 0.3 MeV. The steep rises at around 
$\epsilon_2=0.33$ in Figs.\ref{fig3}(a) and (b) indicate the necessity 
of the $(\pi i_{13/2})^2$ (the [660~1/2] orbital in Fig.\ref{fig2}(c)) alignment 
for the appearance of the wobbling mode although the critical 
value of $\epsilon_2$ itself is frequency dependent. 
Aside from this, $\omega_\mathrm{wob}$ is almost constant in the calculated range. 
The slight increase at around $\epsilon_2=0.4$ stems from the occupation of the 
$\nu$[651~1/2] orbital. We have confirmed that 
in this case the $(\nu j_{15/2})^2$ alignment at around $\epsilon_2=0.47$ seen in 
Fig.\ref{fig3}(b) does not affect $\omega_\mathrm{wob}$ visibly 
since $\mit\Delta\mathcal{J}_y^\mathrm{(eff)}$ in this case is almost the same as 
$\mit\Delta\mathcal{J}_x$ although its reason is not clear. 
Figure~\ref{fig3}(c) graphs $Q_\mathrm{t}$. This figure indicates that the chosen 
shape $\epsilon_2=0.43$ and $\gamma = 20^\circ$ reproduces the measured 
$Q_\mathrm{t}$. 

\begin{figure}[htbp]
  \includegraphics[width=7cm,keepaspectratio]{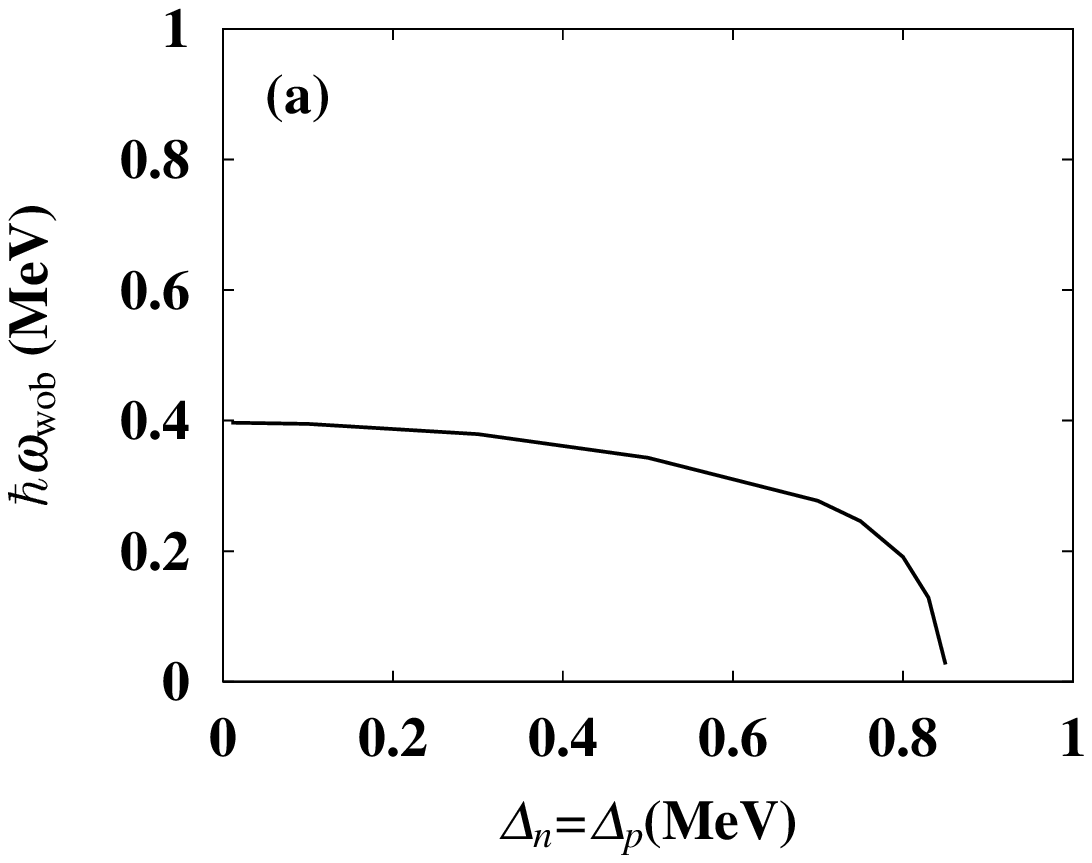}
  \includegraphics[width=7cm,keepaspectratio]{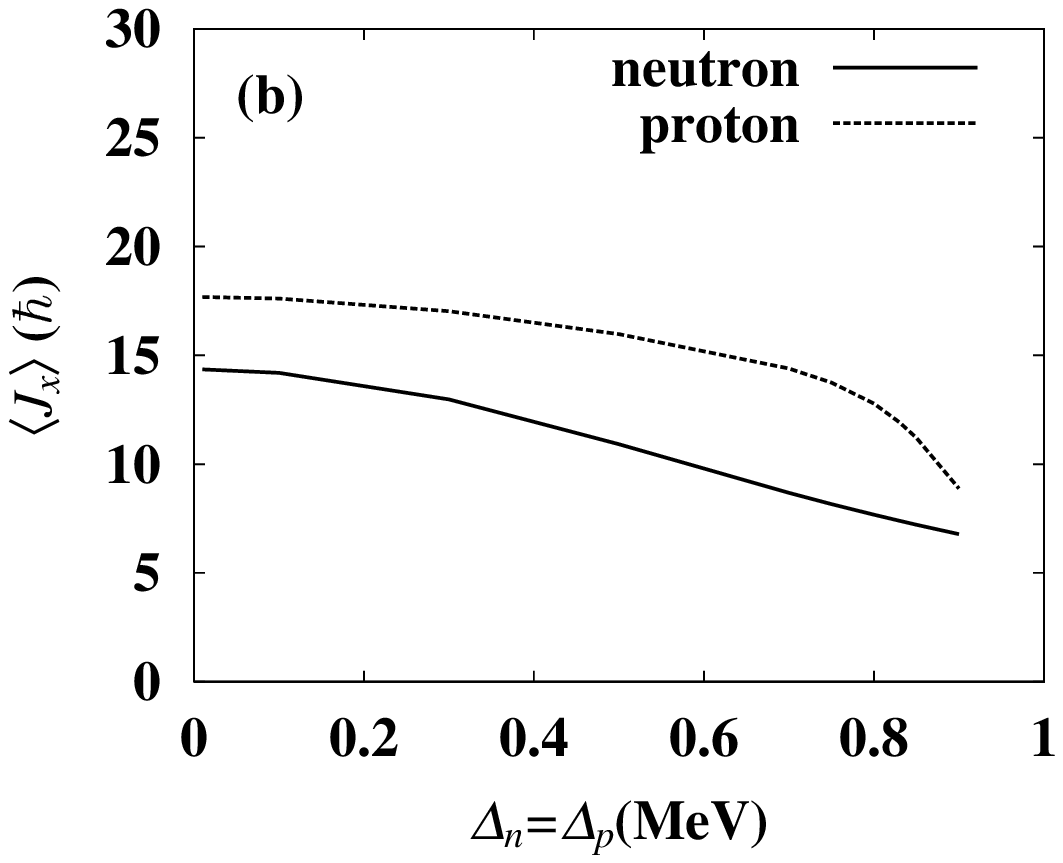}
 \caption{Pairing gap dependence of (a) excitation energy of the wobbling motion 
          and (b) expectation values of angular momenta in the yrast state in $^{168}$Hf, 
          calculated at $\hbar\omega_\mathrm{rot} =$ 0.25 MeV with 
          $\epsilon_2 = 0.43$ and $\gamma = 20^\circ$. 
          $\mit\Delta_n = \mit\Delta_p$ is assumed for simplicity. \label{fig4}}
\end{figure}

 Figure~\ref{fig4}(a) shows dependence on the pairing gaps. Since we do not have 
detailed information about the gaps, we assume $\mit\Delta_n = \mit\Delta_p$ for simplicity. 
This figure shows that the dependence on the gaps is weak unless they are too large. 
Since the static pairing gap $\mit\Delta$ is expected to be small, say, 
$\mit\Delta \le$ 0.6 MeV, in the observed frequency range, 
$\omega_\mathrm{wob}$ is not sensitive to the value of $\mit\Delta$. 
This is a striking contrast to the $\beta$ and $\gamma$ vibrations; it is well 
known that pairing gaps are indispensable for them. Here we note that the 
behavior of the $\omega_\mathrm{wob}$ correlates well with $\langle J_x\rangle_p$ presented in 
Fig.\ref{fig4}(b).

\subsubsection{The odd-$A$ nucleus $^{167}$Lu}

 Next we study $^{167}$Lu in a way similar to the preceding $^{168}$Hf 
case. We choose $\gamma = 20^\circ$ and $\mit\Delta_n = \mit\Delta_p =$ 0.3 MeV as 
representative mean-field parameters as above. As for $\epsilon_2$, however, we 
examined various possibilities because $Q_\mathrm{t}$ has not been measured in this 
nucleus. 
Since the sensitive $\epsilon_2$-dependence through the occupation of the 
$\nu$[651~1/2] orbital appears only at $\hbar\omega_\mathrm{rot} >$ 0.4 MeV and therefore 
the ``band-head" properties do not depend on $\epsilon_2$ qualitatively, first we 
discuss them adopting $\epsilon_2 = 0.43$ in order to look at the difference 
between the even-even and the odd-$Z$ cases. 

\begin{figure}[htbp]
  \includegraphics[width=7cm,keepaspectratio]{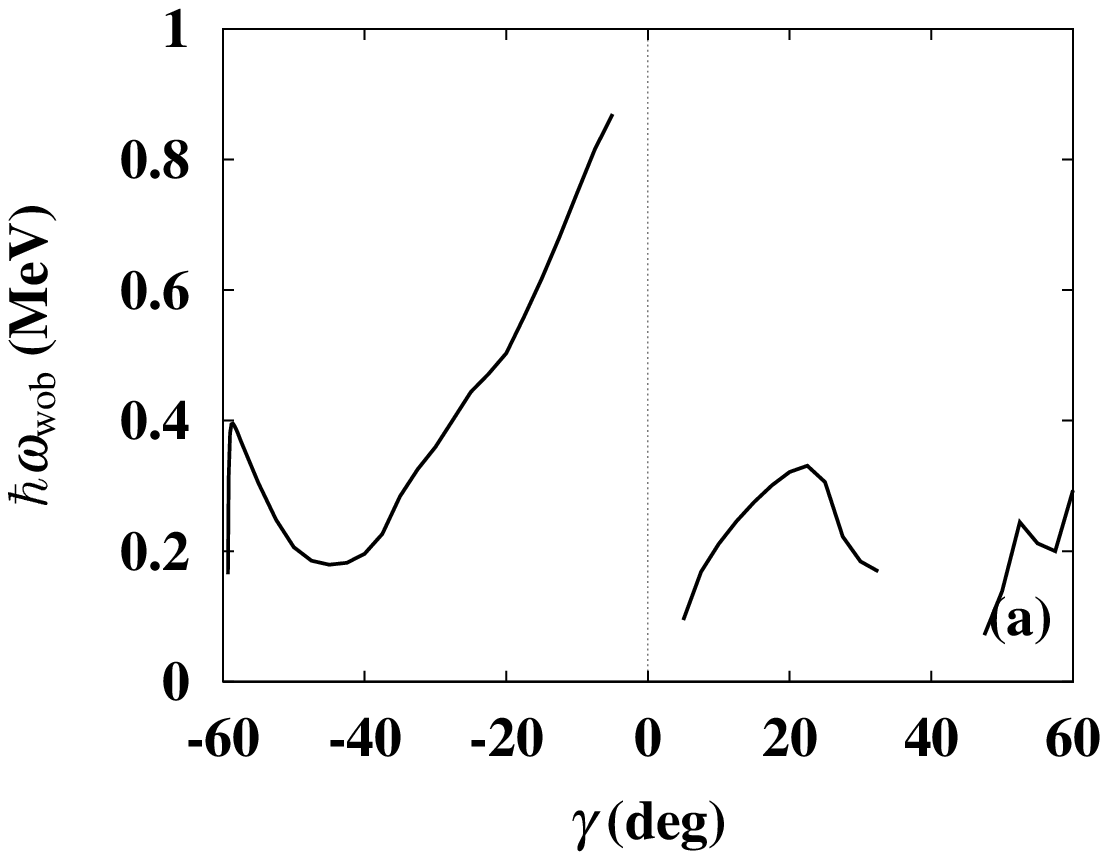}
  \includegraphics[width=7cm,keepaspectratio]{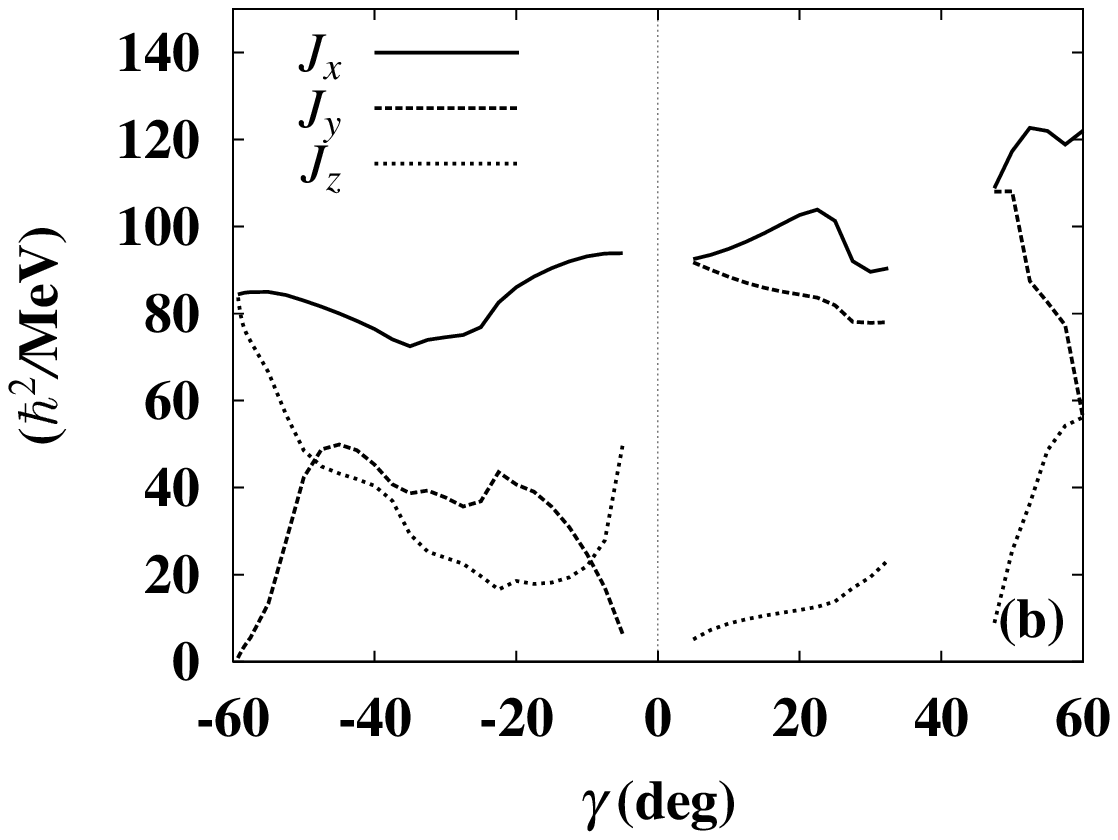}
  \includegraphics[width=7cm,keepaspectratio]{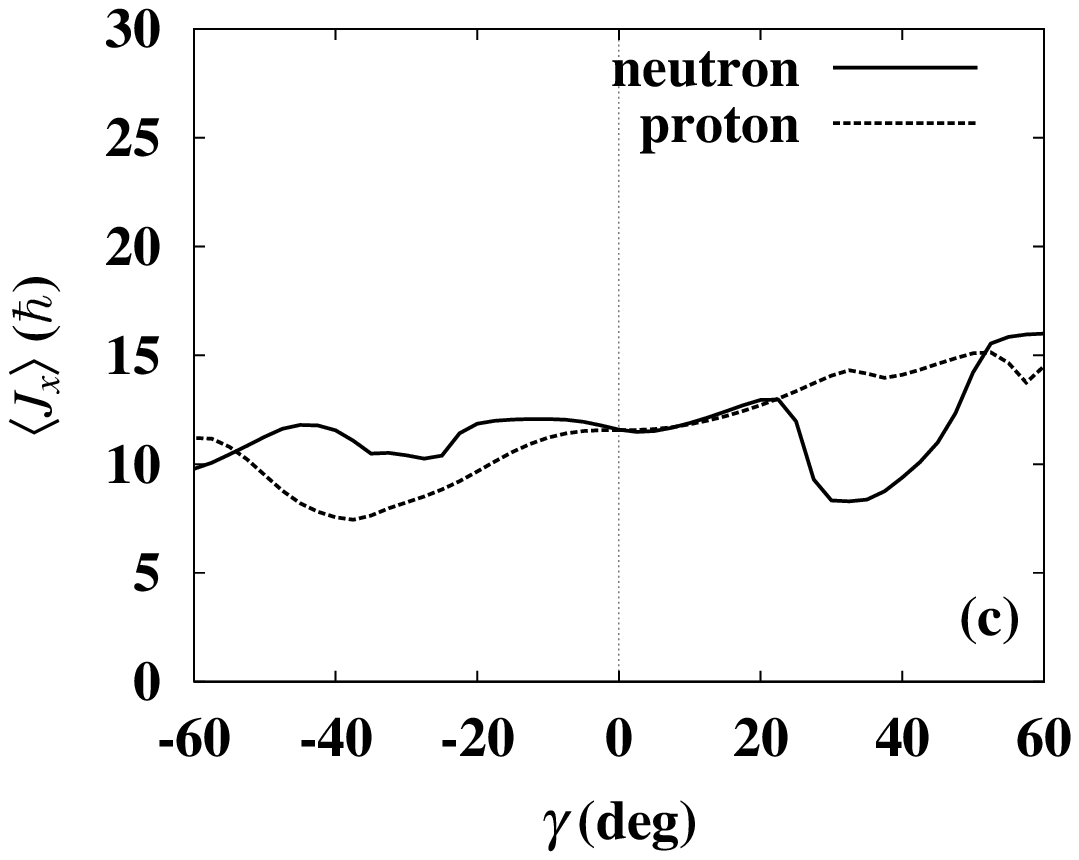}
  \includegraphics[width=7cm,keepaspectratio]{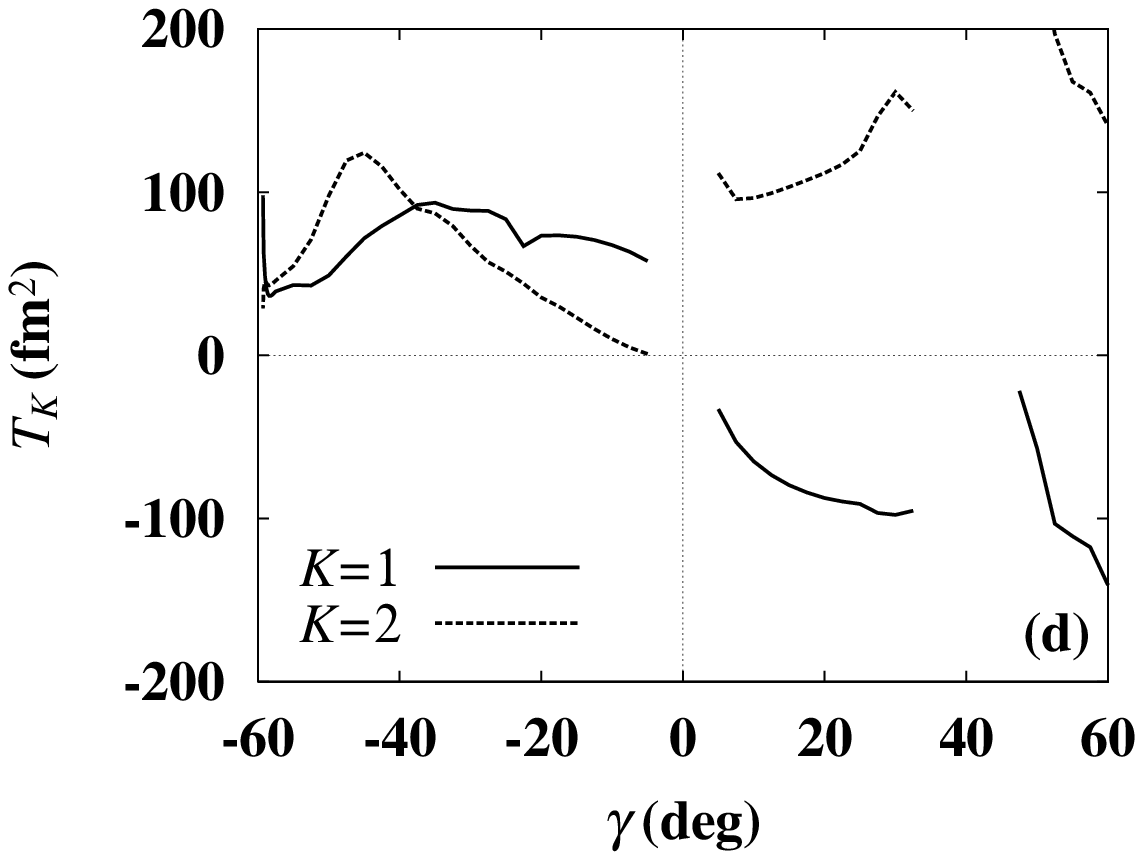}
 \caption{The same as Fig.\ref{fig1} but for $^{167}$Lu. \label{fig5}}
\end{figure}

 Figure~\ref{fig5} shows dependence on $\gamma$ at 
$\hbar\omega_\mathrm{rot} =$ 0.25 MeV with keeping $\epsilon_2 = 0.43$ and 
$\mit\Delta_n = \mit\Delta_p = $ 0.3 MeV constant. Figure~\ref{fig5}(a) graphs $\omega_\mathrm{wob}$. 
In the $\gamma > 0$ region, the solution is quite similar to the $^{168}$Hf case. 
In the $\gamma < 0$ region, that for $-60^\circ \le \gamma \alt -30^\circ$ is 
quite similar again but for $-30^\circ \alt \gamma < 0$ its character is 
completely different. In this region the presented solution is the lowest in  energy 
and becoming collective gradually as $\gamma$ decreases. The largeness of 
$\omega_\mathrm{wob}$ corresponds to that of 
$\mathcal{J}_x-\mathcal{J}_y^\mathrm{(eff)}$ in Fig.\ref{fig5}(b). 
Comparison of Figs.\ref{fig5}(c) and \ref{fig1}(c) certifies that the alignment of 
the $\pi i_{13/2}$ quasiparticle(s) is almost complete for $\gamma > 0$ whereas less 
for $\gamma < 0$. This produces quantitative even-odd differences as explained below. 

\begin{figure}[htbp]
  \includegraphics[width=7cm,keepaspectratio]{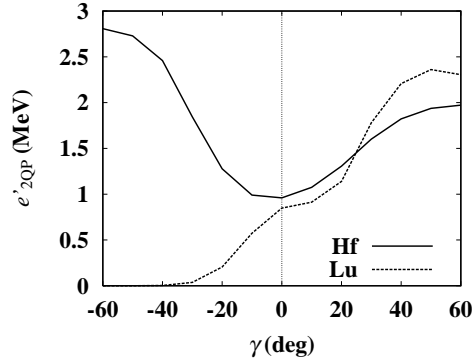}
 \caption{Energies of the lowest $(\pi(N_\mathrm{osc}=6))^2$ two quasiparticle states 
          in $^{168}$Hf and $^{167}$Lu, calculated at the same time as 
          Figs.\ref{fig1} and \ref{fig5}, respectively. \label{fig6}}
\end{figure}

 Having confirmed that these features are independent of $\epsilon_2$ and $N$ 
except that we did not obtain any low-lying solutions for 
$35^\circ \alt \gamma \le 60^\circ$ in the small-$\epsilon_2$ cases, we look into 
underlying unperturbed 2QP energies to see the even-odd difference. 
In Fig.\ref{fig6} we present the energies of the lowest $(\pi(N_\mathrm{osc}=6))^2$ 
states which represent the biggest difference. In the yrast $(\pi i_{13/2})^2$ 
configuration, $A_p$ and $B_p$ in the usual notation are occupied in the 
even-$Z$ case, the lowest 2QP state of signature $\alpha = 1$ with respect to 
this is $\bar B_p C_p$ (where $\bar~$ denotes the conjugate state, see 
Eq.(\ref{exch})). In the odd-$Z$ case in which $A_p$ is occupied, 
the lowest one is $B_p \bar A_p$. 
Since both $e'_{B_p}$ and $e'_{\bar A_p}$ decrease as $\gamma$ decreases, 
this 2QP state becomes the dominant component in the lowest-energy RPA solution. 
Note here that the sum $e'_{B_p}+ e'_{\bar A_p}$ corresponds to the signature 
splitting between $A_p$ and $B_p$ when they are seen from the usual even-even 
vacuum. Since both $B_p$ and $\bar A_p$ are of $K = 1/2$ character, 
the resulting RPA solution can not have the $K = 2$ collectivity as shown in 
Fig.\ref{fig5}(d). According to the relation~\cite{smm},
\begin{equation}
  \frac{\mathcal{J}_y^\mathrm{(eff)}}{\mathcal{J}_x}
 =\left[1 + \frac{\omega_\mathrm{wob}}{\omega_\mathrm{rot}}
                   \frac{\sin{\gamma}}{\sin{(\gamma+\frac{4}{3}\pi)}}
                  \frac{T_1}{T_2} \right]^{-1} ,
\end{equation}
$\mathcal{J}_y^\mathrm{(eff)}$ in Fig.\ref{fig5}(b) becomes small for 
$-30^\circ \alt \gamma < 0$. These discussion serves to exclude the possibility of 
$\gamma \simeq -20^\circ$ for the TSDs that support collective wobbling excitations 
in the odd-$Z$ cases, whereas the even-odd difference in $\gamma > 0$ is merely 
quantitative. 

\subsection{Dependence on the rotational frequency $\omega_\mathrm{rot}$}
\label{secthigh}

\subsubsection{$^{168}$Hf and $^{174}$Hf}

\begin{figure}[htbp]
  \includegraphics[width=7cm,keepaspectratio]{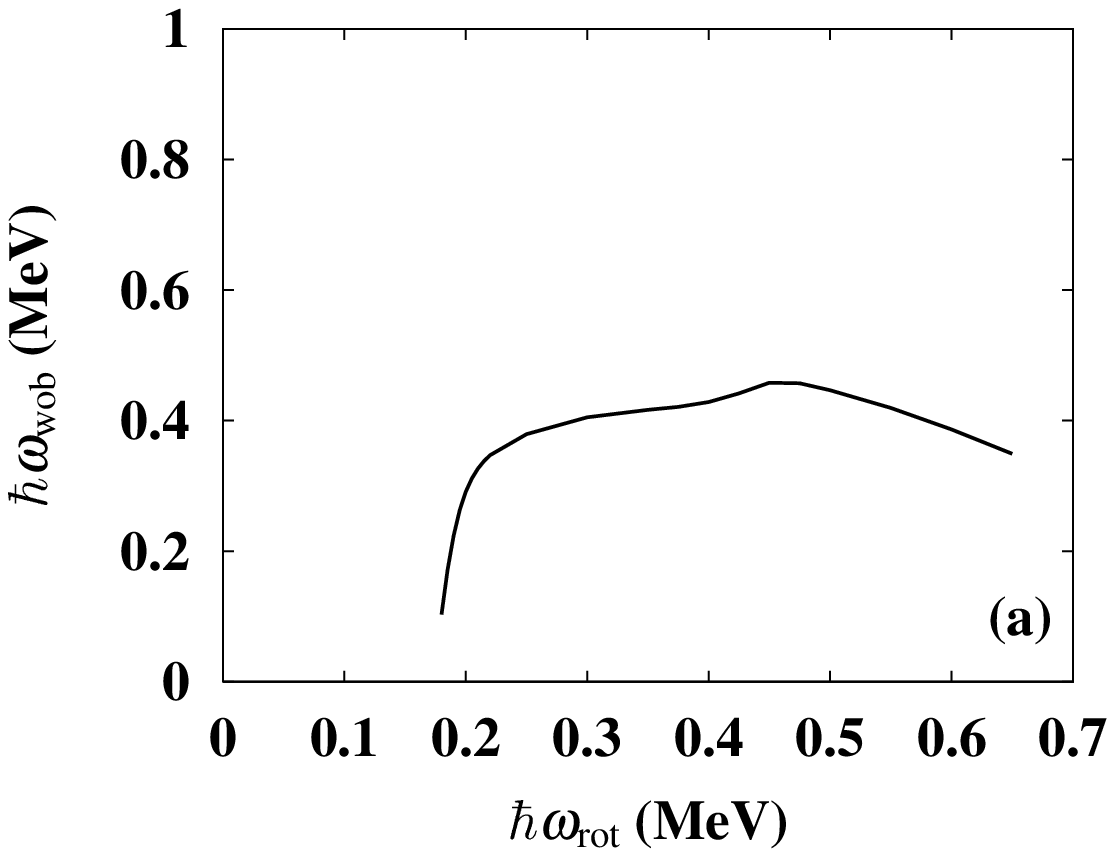}
  \includegraphics[width=7cm,keepaspectratio]{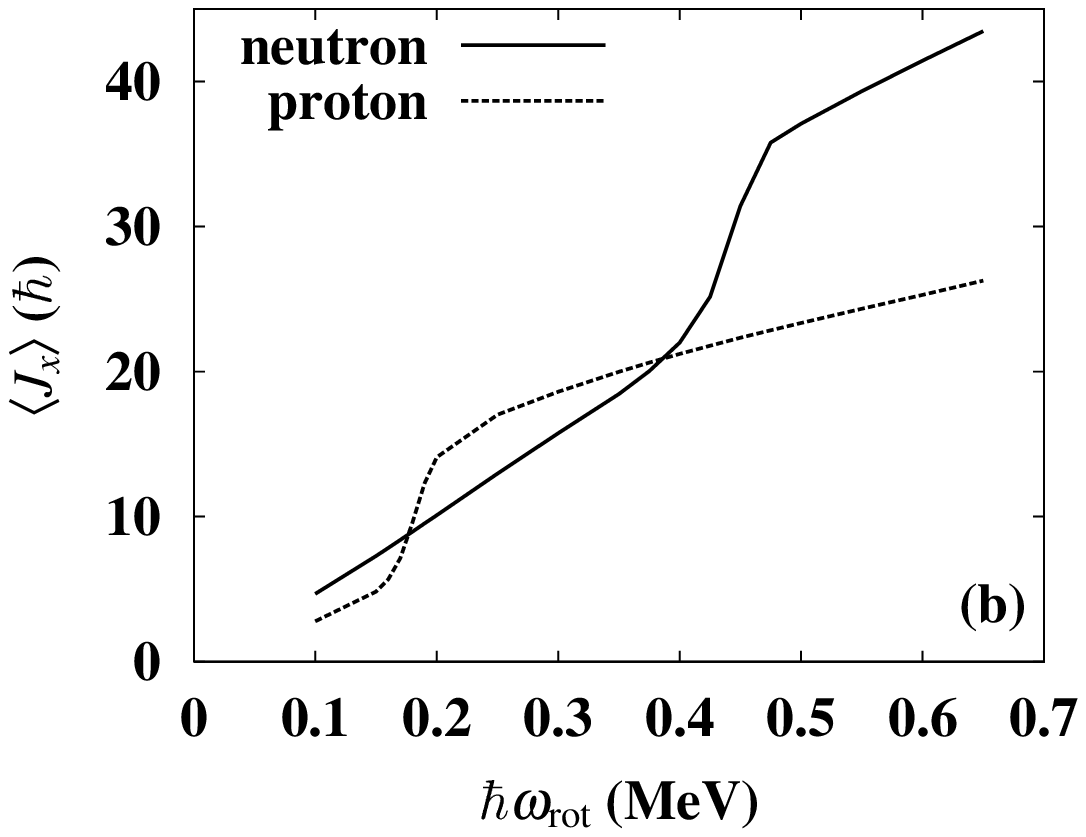}
  \includegraphics[width=7cm,keepaspectratio]{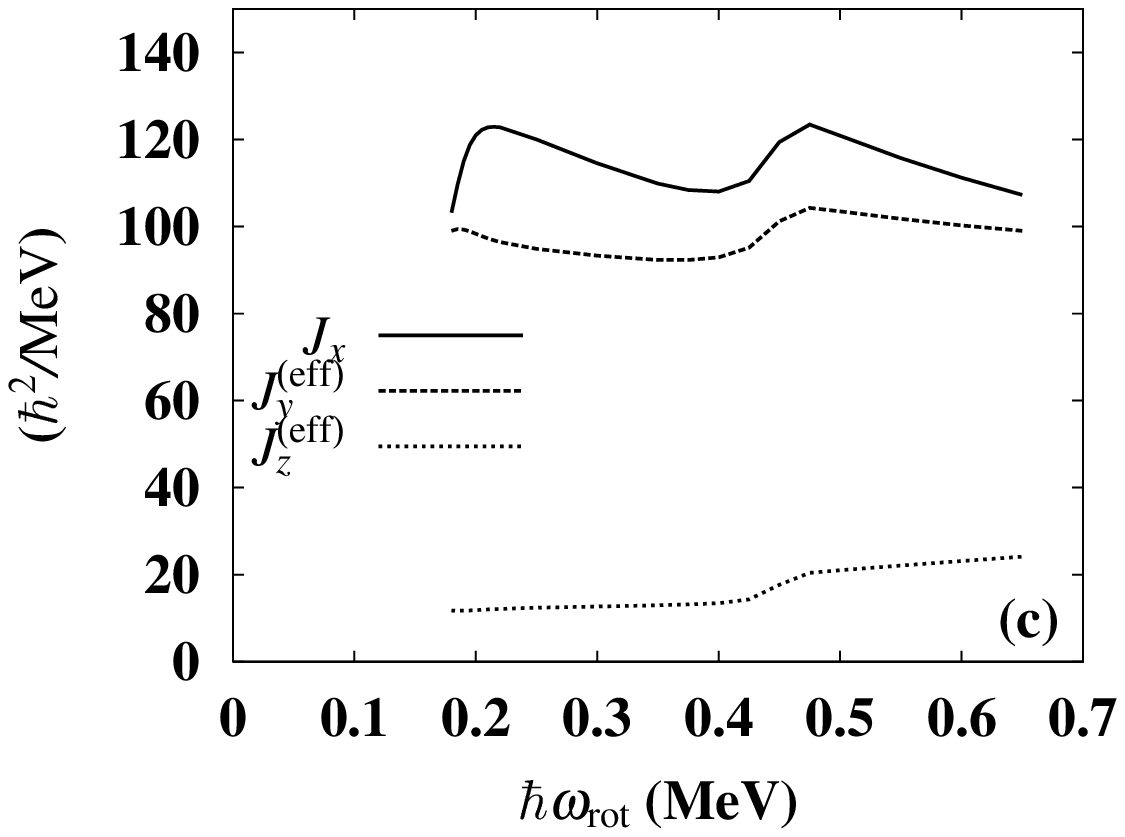}
 \caption{Rotational frequency dependence of (a) excitation energy of the wobbling motion, 
          (b) expectation values of angular momenta in the yrast state, 
          and (c) three moments of inertia associated with the wobbling motion in $^{168}$Hf, 
          calculated with $\epsilon_2 = 0.43$, $\gamma = 20^\circ$ and 
          $\mit\Delta_n = \mit\Delta_p =$ 0.3 MeV. \label{fig7}}
\end{figure}

 The analyses above indicate that the chosen mean-field parameters are 
reasonable, and therefore we proceed to study $\omega_\mathrm{rot}$-dependence 
with keeping these parameters constant. Figure~\ref{fig7} shows the result for 
$^{168}$Hf. 
These figures indicate again the $(\pi i_{13/2})^2$ alignment that makes 
$\mathcal{J}_x$ larger than $\mathcal{J}_y^\mathrm{(eff)}$ is indispensable 
for the formation of the wobbling excitation. At around 
$\hbar\omega_\mathrm{rot} =$ 0.45 MeV the $(\nu j_{15/2})^2$ alignment occurs. 
In contrast to the low-frequency case reported in Fig.\ref{fig3}, in the present case 
its effect on $\omega_\mathrm{wob}$ is visible as a small bump. 
Although the character of the observed excited TSD bands has not been resolved, 
some anomaly is seen at around this $\omega_\mathrm{rot}$ in one of them~\cite{hf168}. 
We suggest this is related to the $(\nu j_{15/2})^2$ alignment since this is the only 
alignable orbital in this frequency region of this shape. 
However we note that in $^{167}$Lu an interaction with a normal deformed 
state at around this frequency is discussed in Ref.\cite{lu167}. 

 We performed calculations also for $\gamma = -20^\circ$. In that case, however, 
wobbling excitation exists only at small $\omega_\mathrm{rot}$ because 
$\mathcal{J}_x-\mathcal{J}_y^\mathrm{(eff)}$ is small as seen from Fig.\ref{fig1}(b). 

\begin{figure}[htbp]
  \includegraphics[width=7cm,keepaspectratio]{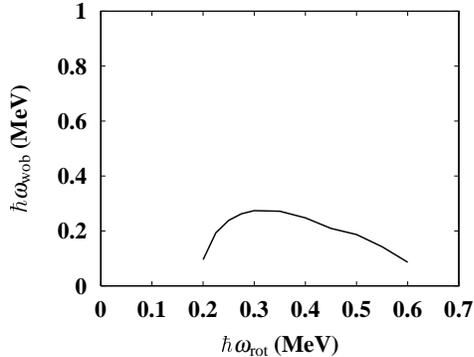}
 \caption{Rotational frequency dependence of excitation energy of the wobbling motion 
          in $^{174}$Hf, calculated with $\epsilon_2 = 0.453$, $\gamma = 16^\circ$ and 
          $\mit\Delta_n = \mit\Delta_p =$ 0.3 MeV. \label{fig8}}
\end{figure}

 Very recently TSD bands were observed in another even-even nucleus, 
$^{174}$Hf~\cite{hf174}. It is not trivial if a similar band structure is 
observed in the nucleus with six neutrons more since the existence of the TSD states 
depends on the shell gap. Multiple TSD bands were observed but connecting 
$\gamma$-rays have not been resolved also in this nucleus. 
We performed a calculation adopting $\epsilon_2 = 0.453$ and $\gamma = 16^\circ$ 
suggested in Ref.\cite{hf174} and $\mit\Delta_n = \mit\Delta_p =$ 0.3 MeV. 
The result is presented in Fig.\ref{fig8}. 
The most striking difference from the case of $^{168}$Hf above is that 
$\omega_\mathrm{wob}$ decreases steadily as $\omega_\mathrm{rot}$ increases 
after the $(\pi i_{13/2})^2$ alignment is completed. This is because the 
$(\nu j_{15/2})^2$ alignment that causes the small bump in the $^{168}$Hf case shifts 
to very low $\omega_\mathrm{rot}$ due to the larger neutron number. 

\subsubsection{$^{167}$Lu}

 The wobbling excitation was first observed experimentally in $^{163}$Lu~\cite{lu1}, 
later it was also observed in $^{165}$Lu~\cite{lu165} and $^{167}$Lu~\cite{lu167}. 
The characteristic features common to these isotopes are 1) $\omega_\mathrm{wob}$ 
decreases as $\omega_\mathrm{rot}$ increases contrary to the consequence of 
calculations adopting constant moments of inertia, and 2) 
$B(E2:I \rightarrow I-1)_\mathrm{out} / B(E2:I \rightarrow I-2)_\mathrm{in}$ 
is large --- typically around 0.2. 

\begin{figure}[htbp]
  \includegraphics[width=7cm,keepaspectratio]{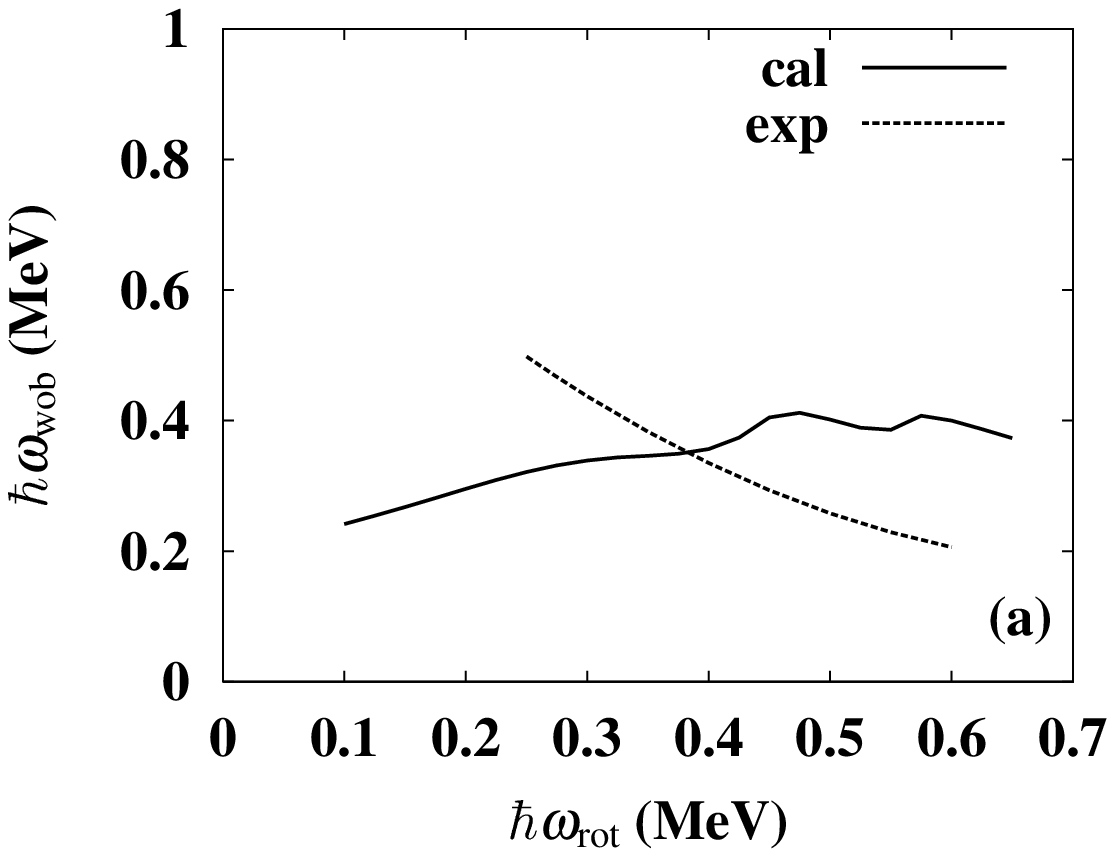}
  \includegraphics[width=7cm,keepaspectratio]{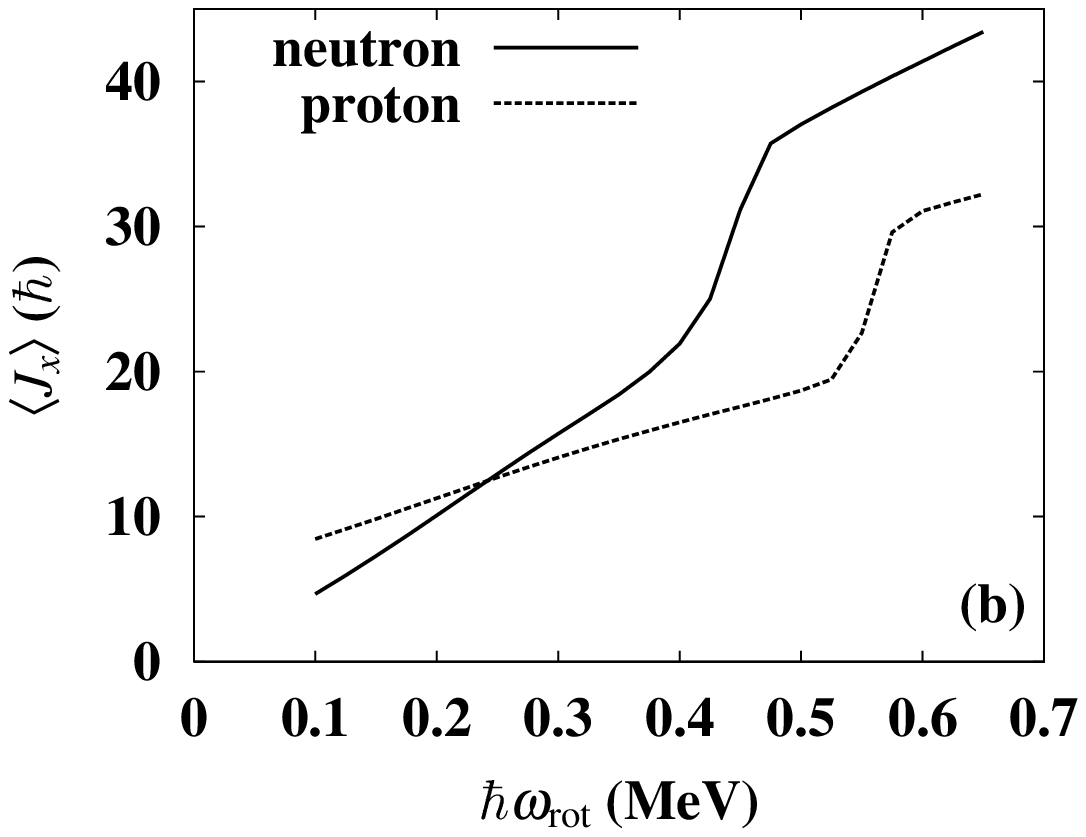}
  \includegraphics[width=7cm,keepaspectratio]{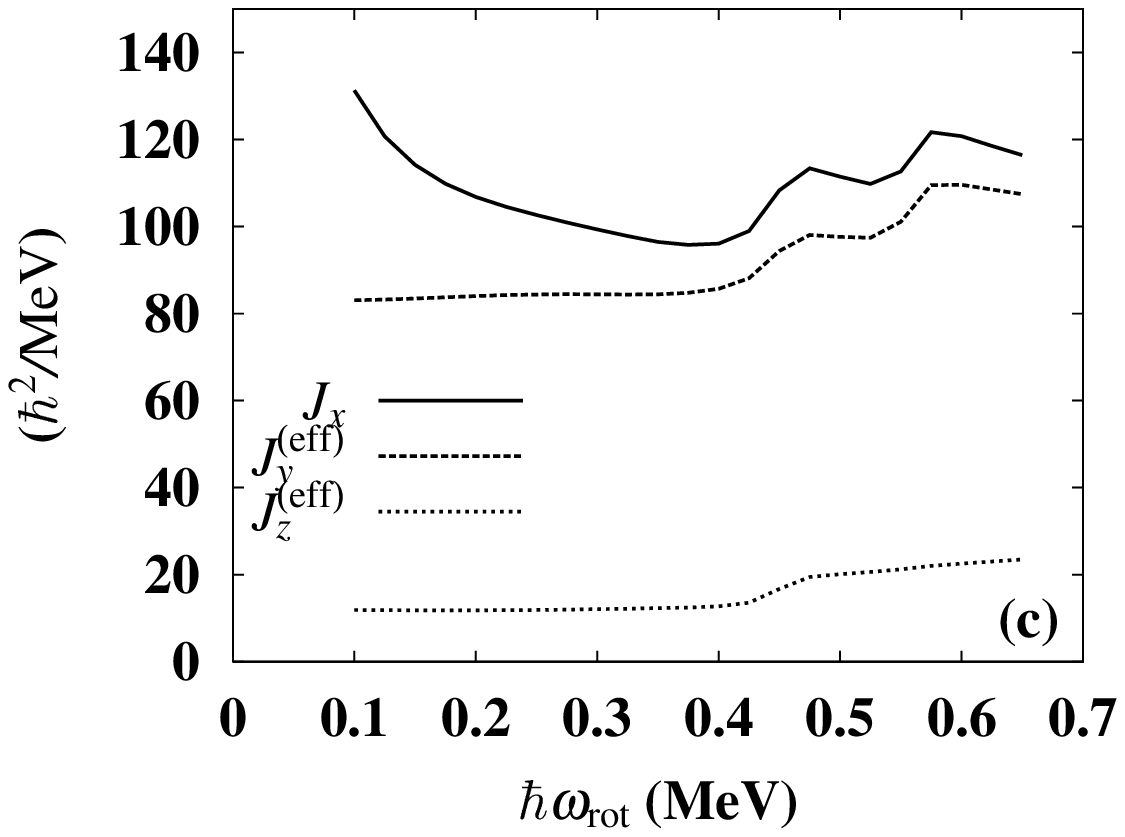}
 \caption{The same as Fig.\ref{fig7} but for $^{167}$Lu. Experimental values taken 
          from Ref.\cite{lu167} are also included in (a). \label{fig9}}
\end{figure}

 Here we concentrate on the isotone of $^{168}$Hf discussed above, that is, $^{167}$Lu
in order to see the even-odd difference. A comparison of Figs.\ref{fig7} and \ref{fig9} 
proves that all the differences are due to the fact that the number of the aligned 
$\pi i_{13/2}$ quasiparticle is less by one: 1) The $(\pi i_{13/2})^2$ alignment at around 
$\hbar\omega_\mathrm{rot}$ = 0.2 MeV is absent, and 2) the $B_pC_p$ crossing occurs at around 
$\hbar\omega_\mathrm{rot}$ = 0.55 MeV, which is proper to the $(\pi i_{13/2})^1$ configuration. 
Figure~\ref{fig9}(a) shows that our calculation does not reproduce the data, although in 
each frequency range in which the configuration is the same $\omega_\mathrm{wob}$ decreases 
at high $\omega_\mathrm{rot}$ as in the cases of the even-even nuclei presented above. 
This result might indicate that there is room for improving the mean field.
The $\mathcal{J}_x$ in Fig.\ref{fig9}(c) is larger than the experimentally deduced value  by 
about 20 -- 30 \%. This is due to the spurious velocity dependence of the Nilsson potential 
mentioned in Sec.\ref{sectwob}.

\subsection{Interband $B(E2)$ transitions}
\label{sectbe2}

\begin{figure}[htbp]
  \includegraphics[width=7cm,keepaspectratio]{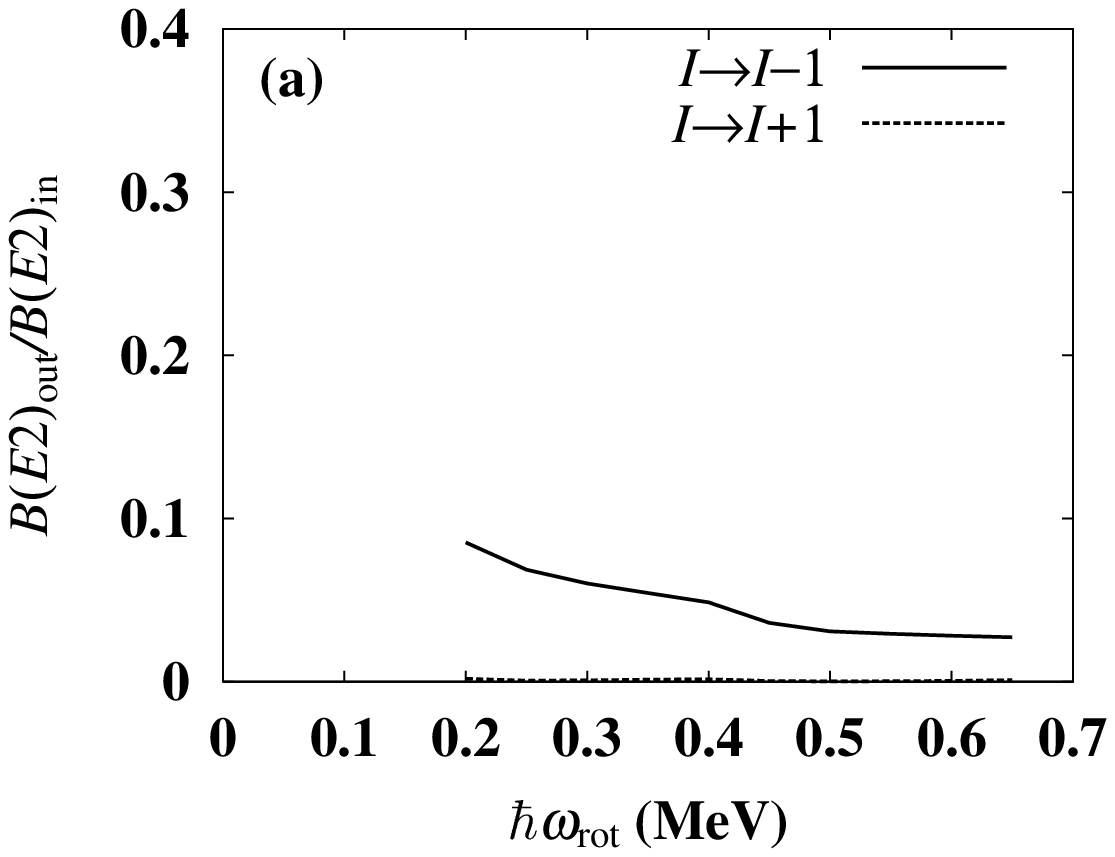}
  \includegraphics[width=7cm,keepaspectratio]{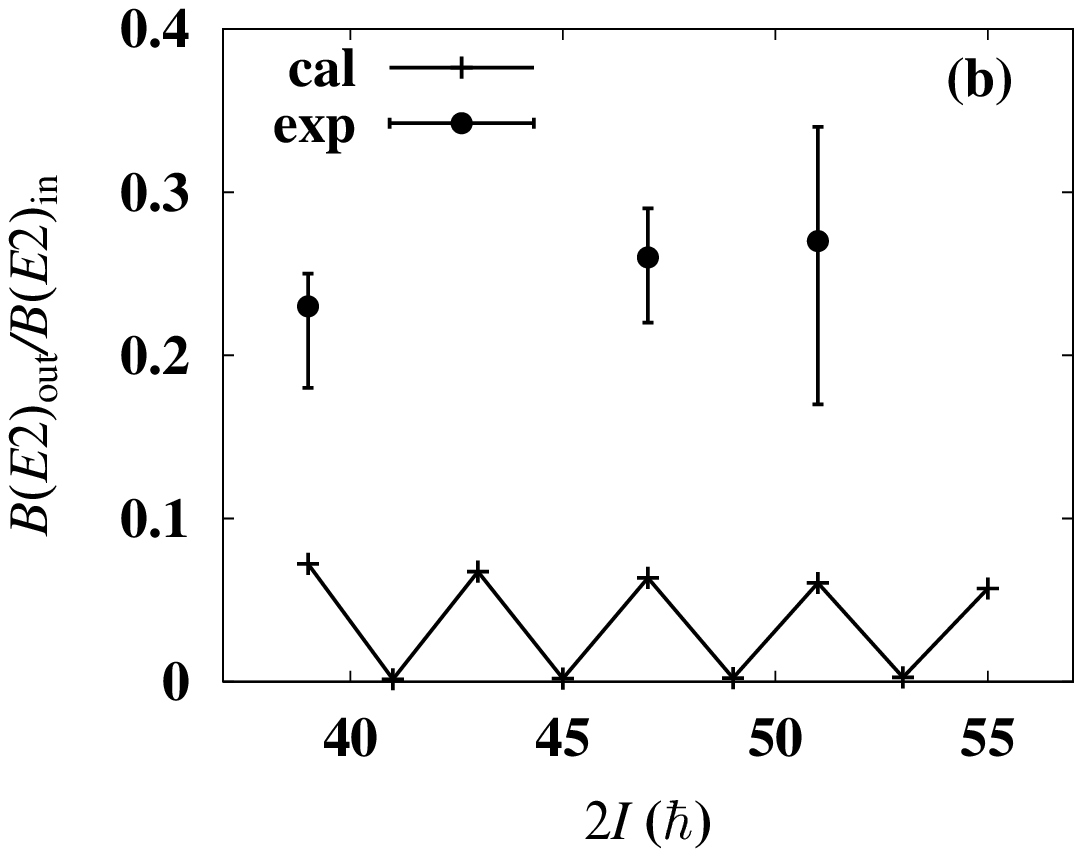}
 \caption{Interband $E2$ transition rates for 
          $I$ (wobbling on yrast TSD) $\rightarrow$ $I\pm1$ (yrast TSD) transitions in (a) $^{168}$Hf 
          and (b) $^{167}$Lu. The latter is presented as functions of $2\times$ spin $I$, 
          while the former is presented as functions of the rotational frequency since 
          experimental spin assignment has not been done for $^{168}$Hf. 
          The rotational-frequency range corresponding to (b) is very narrow in comparison to (a).
          Interband transition rates are divided by the in-band ones. 
          Experimental values~\cite{lu167} are also shown in (b). 
          Noting that, for $^{167}$Lu, the states $I+1$ (TSD1) are slightly higher 
          in energy than $I$ (TSD2) at $I>51/2$$\hbar$ and 
          $B(T_\lambda;I\rightarrow I+1)\simeq B(T_\lambda;I+1\rightarrow I)$ holds 
          at high spins, we plotted those for $I\rightarrow I+1$ at the places with the 
          abscissae $I+1$ in order to show clearly their characteristic staggering 
          behavior.  \label{fig10}}
\end{figure}

  Compared to the excitation energy, the interband $B(E2)$ values relative
to the in-band ones have been measured in only few cases.
In Fig.\ref{fig10}, we report calculated $B(E2)$ ratios for
$I$ (wobbling on yrast TSD) $\rightarrow$ $I \pm 1$ (yrast TSD) transitions 
in $^{168}$Hf and $^{167}$Lu.
The measured ones are also included for the latter. 

 The first point is the magnitude of the larger ($I\rightarrow I-1$) ones. 
Apparently, the calculated $B(E2)$ values are smaller by factor 2 -- 3. 
The measured interband $B(E2)$ values amount almost to
the macroscopic rotor value.  In the RPA calculations,
as summarized in sect.\ref{sectwob}, the $B(E2)$ value is reduced
by a factor $c_{n=\mathrm{wob}}^2$ (see Eq.~(\ref{BE2wob})):
Only the case with the full-strength
$c_{n=\mathrm{wob}}^2=1$ the rotor value is recovered.
Although the obtained RPA wobbling solutions are extremely collective
in comparison with the usual low-lying collective vibrations,
like the $\beta$- or $\gamma$-vibrations, for which typically
$|c_n| \simeq$ 0.3 -- 0.4, this factor is still
$|c_{n=\mathrm{wob}}| \simeq$ 0.6 -- 0.8.  This is the main reason why
the calculated $B(E2)$ values are a factor 2 -- 3 off the measured ones.
As is well-known, giant resonances also carry considerable
amount of quadrupole strengths, so it seems difficult for the microscopic
correction factor $c_{n=\mathrm{wob}}^2$ to be unity; it is not impossible,
however, because the ``sum rule'' discussed in sect.\ref{sectwob}
is not the sum of positive-definite terms.
In the RPA formalism, the reduction factor $c_{n=\mathrm{wob}}^2$
for the $B(E2)$ value, Eq.~(\ref{BE2wob}), comes from
the fact that the wobbling motion is composed of
the coherent motion of two-quasiparticles, and
reflects the microscopic structure of collective RPA solutions.
The measurement that the $B(E2)$ value suffers almost no reduction may
be a challenge to the microscopic RPA theory in the case of
the wobbling motion. 
Calculated $B(E2)$ ratios for $^{174}$Hf are slightly smaller than those for 
$^{168}$Hf in Fig.\ref{fig10}(a). 

 The second point is the staggering, that is, the difference between 
$I\rightarrow I \pm 1$. We clarified~\cite{smm} its unique correspondence to the sign of 
$\gamma$ as mentioned in subsect.\ref{sectlow}; that holds for both even-even and odd-$A$ 
systems. Recently this staggering was discussed from a different point of view~\cite{casten}; 
but it looks to apply only to $\gamma<0$ cases. 

\section{Conclusion}

 The nuclear wobbling motion, which is a firm evidence of stable triaxial deformations, 
was identified experimentally in the triaxial superdeformed odd-$A$ Lu isotopes. 
In principle, wobbling excitation is possible both in $\gamma>0$ and $\gamma<0$ nuclei. 
Every information, theoretical and experimental, suggests $\gamma>0$ for these bands. 
According to the wobbling frequency formula~\cite{bm}, c.f. Eq.(\ref{disp}),
its excitation in nuclei rotating principally about the $x$ axis requires 
$\mathcal{J}_x > \mathcal{J}_y, \mathcal{J}_z$, although irrotational-like model 
moments of inertia give $\mathcal{J}_x < \mathcal{J}_y$ for $\gamma>0$. 
To solve this puzzle, we studied the nuclear wobbling motion, in particular, 
the three moments of inertia associated with it in terms of the cranked shell 
model plus random phase approximation. 
This makes it possible to calculate the moments of inertia of the whole system 
including the effect of aligned quasiparticle(s). The results indicate that the 
$\gamma$-dependence of the calculated moment of inertia is basically irrotational-like 
($\mathcal{J}_x \gtrless \mathcal{J}_y$ for $\gamma \lessgtr 0$) if aligned 
quasiparticle(s) ($\pi i_{13/2}$ in the present case) does not exist. 
But once it is excited, it produces an additional contribution, 
$\mit\Delta\mathcal{J}_x=i_\mathrm{QP}/\omega_\mathrm{rot}$, and consequently can lead 
to $\mathcal{J}_x > \mathcal{J}_y$. This is the very reason why wobbling excitation 
exists in $\gamma>0$ nuclei. In this sense, the wobbling motion is a collective motion 
that is sensitive to the single-particle alignments. 

The resulting moment of inertia for $0 <\gamma \alt 30^\circ$ resembles the $\gamma$-reversed 
one, i.e., the irrotational moment of inertia
but with $\mathcal{J}_x$ and $\mathcal{J}_y$ being interchanged.
That for $50^\circ \alt \gamma \alt 60^\circ$, where single-particle angular 
momenta dominate, is rigid-body-like. 
That for $\gamma<0$ is irrotational-like except for odd-$A$ nuclei with 
$-30^\circ \alt \gamma < 0$ where a specific 2QP state determines the lowest RPA 
solution. 

 Having studied qualitative features of the three moments of inertia at a low rotational 
frequency, we calculated wobbling bands up to high $\omega_\mathrm{rot}$. 
Experimentally they were confirmed only in odd-$A$ Lu isotopes as mentioned above. 
The most characteristic feature of the data is that $\omega_\mathrm{wob}$ 
decreases as $\omega_\mathrm{rot}$ increases. This obviously excludes constant 
moments of inertia. In our calculation three moments of inertia are automatically 
$\omega_\mathrm{rot}$-dependent even when mean field parameters are fixed constant. 
It should be stressed that the wobbling-like solution in our RPA calculations
is insensitive to the mean-field parameters, especially to the pairing gaps,
as is shown in subsect.\ref{sectlowHf}.
This distinguishes the wobbling-like solution from the usual collective vibrations,
which are sensitive to the pairing correlations.
Thus, our microscopic RPA calculation confirms that the observed band
is associated with a new type of collective excitation, although
comparisons to the observed excitation energy indicate that there is room for 
improving the calculation.

 As for the interband transition rates, 
our calculation accounted for only about one half or less of the measured ones,
even though the wobbling-like solution is extremely collective compared to
the usual vibrational modes.
This issue is independent of the details of choosing parameters. 
This confronts microscopic theories with a big challenge. 

\begin{acknowledgments}
We thank G. B. Hagemann for providing us with some experimental information 
prior to publication. 
This work was supported in part by the Grant-in-Aid for scientific research 
from the Japan Ministry of Education, Science and Culture 
(Nos. 13640281 and 14540269).
\end{acknowledgments}

\end{document}